\documentclass[preprint]{aastex61}
\mathchardef\mhyphen="2D
\shorttitle{MUSUBI}
\shortauthors{Wang et al.}

\begin{document}

\title{MUSUBI (MegaCam Ultra-deep Survey: $u^\ast$-Band Imaging)--Data for the COSMOS and SXDS Fields}

\author{Wei-Hao Wang}
\affiliation{Academia Sinica Institute of Astronomy and Astrophysics, No.\ 1, Sec.\ 4, Roosevelt Rd., Taipei 10617, Taiwan}

\author{Sebastien Foucaud}
\affiliation{Department of Astronomy, Shanghai Jiao Tong University, Shanghai, China}

\author{Bau-Ching Hsieh}
\affiliation{Academia Sinica Institute of Astronomy and Astrophysics, No.\ 1, Sec.\ 4, Roosevelt Rd., Taipei 10617, Taiwan}

\author{Hung-Yu Jian}
\affiliation{Academia Sinica Institute of Astronomy and Astrophysics, No.\ 1, Sec.\ 4, Roosevelt Rd., Taipei 10617, Taiwan}

\author{Lihwai Lin}
\affiliation{Academia Sinica Institute of Astronomy and Astrophysics, No.\ 1, Sec.\ 4, Roosevelt Rd., Taipei 10617, Taiwan}

\author{Yen-Ting Lin}
\affiliation{Academia Sinica Institute of Astronomy and Astrophysics, No.\ 1, Sec.\ 4, Roosevelt Rd., Taipei 10617, Taiwan}

\author{Jean Coupon}
\affiliation{Astronomy Department, University of Geneva, Chemin d'Ecogia 16, CH-1290 Versoix, Switzerland}

\author{Yasuhiro Hashimoto}
\affiliation{Department of Earth Sciences, National Taiwan Normal University No. 88, Section 4, Tingzhou Rd., Wenshan District, Taipei 11677, Taiwan}

\author{Masami Ouchi}
\affiliation{Institute for Cosmic Ray Research, The University of Tokyo, 5-1-5 Kashiwanoha, Kashiwa, Chiba 277-8582, Japan}
\affiliation{National Astronomical Observatory of Japan, 2-21-1 Osawa, Mitaka, Tokyo 181-8588, Japan}
\affiliation{Kavli Institute for the Physics and Mathematics of the Universe (WPI), University of Tokyo, Kashiwa, Chiba 277-8583, Japan}

\author{Kazuhiro Shimasaku}
\affiliation{Department of Astronomy, School of Science, The University of Tokyo, 7-3-1 Hongo, Bunkyo-ku, Tokyo 113-0033, Japan}
\affiliation{Research Center for the Early Universe, The University of Tokyo, 7-3-1 Hongo, Bunkyo-ku, Tokyo 113-0033, Japan}

\author{Youichi Ohyama}
\affiliation{Academia Sinica Institute of Astronomy and Astrophysics, No.\ 1, Sec.\ 4, Roosevelt Rd., Taipei 10617, Taiwan}

\author{Keiichi Umetsu}
\affiliation{Academia Sinica Institute of Astronomy and Astrophysics, No.\ 1, Sec.\ 4, Roosevelt Rd., Taipei 10617, Taiwan}

\author{Shiang-Yu Wang}
\affiliation{Academia Sinica Institute of Astronomy and Astrophysics, No.\ 1, Sec.\ 4, Roosevelt Rd., Taipei 10617, Taiwan}

\author{Tzu-Ching Chang}
\affiliation{Jet Propulsion Laboratory, California Institute of Technology, Pasadena, CA 91101, USA}
\affiliation{California Institute of Technology, Pasadena, CA 91125, USA}
\affiliation{Academia Sinica Institute of Astronomy and Astrophysics, No.\ 1, Sec.\ 4, Roosevelt Rd., Taipei 10617, Taiwan}

\begin{abstract}
The Subaru Hyper Suprime-Cam (HSC) Strategic Survey is the latest-generation multi-band optical imaging survey for galaxy evolution and structure formation. The ``Ultra-Deep'' component of the HSC survey provides $grizy$ broad-band images over $\sim3.4$~deg$^2$ to detection limits of $\sim26$--28 AB, along with narrow-band images, in the COSMOS and the SXDS fields. These images provide an unprecedented combination of depths and area coverage, for the studies galaxies up to $z\sim7$. However, the lack of coverage at $<4000$~\AA ~implies incomplete sampling of the rest-frame UV at $z\lesssim 3$, which is critically needed for understanding the buildup of stellar mass in the later cosmic time. We conducted a multi-year CFHT $u^\ast$-band imaging campaign in the two HSC Ultra-Deep fields with CFHT MegaCam.  By including shallower archival data, we reach 5-$\sigma$ depths of $u^\ast=28.1$ and 28.4 (AB) at the centers of the COSMOS and SXDS fields, respectively, and $u^\ast=27.7$ and 27.8 in the central 1 deg$^2$ fields. The image quality is $\gtrsim 0 \farcs90$, fairly good for the $u^\ast$ band. Both the photometric and astrometric quality of our data are excellent. We show that the combination of our $u^\ast$-band and HSC data can lead to high-quality photometric redshifts at $z=0$--3, and robust measurements of rest-frame UV on galaxies at $0.4<z<0.6$ for distinguishing green-valley galaxies from star-forming and quiescent galaxies.  We publicly release our reduced $u^\ast$-band images and reference catalogs that can be used readily for scientific studies.

\end{abstract}


\section{Introduction}

Ever since the SDSS survey \citep{yor00} two decades ago, optical imaging and spectroscopic surveys have revolutionized our views on the galaxy evolution and structure formation. In particular, multi-band imaging surveys over large areas, such as PanSTARRS \citep{kai10}, Dark Energy Survey \citep{sev21}, Hyper Suprime-Cam Subaru Strategic Program \citep[HSC SSP,][]{aihara18}, and the forthcoming LSST \citep{ive19}, have become a standard and relatively inexpensive pathway in mapping the distribution of galaxies over the cosmic time by utilizing the so-called photometric redshift (hereafter photo-$z$) technique.

The HSC SSP is a 7-year program conducted during 2014-2021 on the 8.2-m Subaru Telescope, consisting of different combinations of depths and areas (Wide, Deep, and Ultra-Deep). The detector is optimized to observe in the redder optical wavelengths. Five broad-band ($g, r, i, z$ and $y$) and four narrow-band ($NB387, NB816, NB921, NB101$) filters are designed to make the best use of its sensitivity. Among the three HSC survey layers, the HSC Ultra-Deep Survey (UDS), is the deepest component, aiming to reach 5-$\sigma$ AB magnitude detection limits of $g$, $r$, $i$ $\sim28.0$, $z\sim 27.0$, $y\sim26$, and $m_{NB}\sim26$ over two fields widely separated on the sky (each with one HSC pointing, which covers $\sim 1.7$~deg$^{2}$), making it the deepest survey with a few square-degree of coverage ever taken by a ground-based telescope. For comparisons, these correspond to the depths of Great Observatories Origins Deep Survey \citep[GOODS,][]{gia04} but for an area that is 40$\times$ larger, and to depths of 1--2 magnitudes deeper than the Subaru Suprime-Cam data in the original Cosmic Evolution Survey \citep[COSMOS,][]{sco07,cap07} with almost twice the area. The goal the HSC UDS is to directly measure the buildup of galaxies and large scale structure across cosmic time. 

The two pointings chosen for the HSC UDS are two extremely data-rich extragalactic survey fields, the COSMOS field and the Subaru-XMM Deep Survey \citep[SXDS,][]{fur08} field. As we would like to measure the luminosity and mass functions accurately, sampling a representative volume of the Universe is crucial. One HSC pointing will map 150$^{2}$ comoving Mpc$^{2}$ at $z\sim3$. However, cosmic variance in one pointing will be at a level of 5-15\% depending on the redshift. By observing two independent fields, we can bring it down to 4-10\% \citep{dri10}. 

While the HSC UDS is expected to discover several millions of galaxies up to $z \sim7$, it does not sample the wavelengths blueward of 4000 \AA, which is critical to distinguish the Balmer and Lyman breaks for the precise measurements of photo-$z$, as well as to provide UV-based star formation rate for galaxies at intermediate redshifts. To complement the HSC UDS, we initiated a multi-year $u^\ast$-band imaging campaign in the two HSC UDS fields using MegaCam on Canada-France-Hawaii Telescope, called ``MegaCam Ultra-Deep Survey: $u^\ast$-Band Imaging'' (MUSUBI). This takes advantage of the good $u^\ast$-band quantum efficiency of the MegaCam CCD (74\% at 3800 \AA, cf. 36\% for the HSC CCD at 3800 \AA), which makes MegaCam the best instrument worldwide for this kind of ultraviolet surveys. By combining MUSUBI and existing shallower $u^\ast$-band observations in COSMOS and SXDS, we reach a depth of $u^\ast_{AB}$ $\sim$ 27.5, which is well-matched to the HSC $grizy$ and NB depth. The combination of the MUSUBI and HSC datasets will enable a variety of scientific studies, such as  studies of Lyman-alpha emitters \citep{hu02,ouc08,shi18} and UV-luminosity function for galaxies at $2 < z < 3$ \citep{red09,saw12,van10,mou20},  properties of LBG/BM/BX selected populations \citep{ade04,ly12}, and selection of green-valley galaxies at $z < 1$ using the $NUV-r$ color \citep{wyd07,sal14,coe18}

From 2014, another CFHT MegaCam $u$/$u^\ast$-band imaging campaign was launched by CLAUDS (CFHT Large Area $U$-band Deep Survey, \citealp{sawicki19}).  CLAUDS imaged the four  ``Deep Fields'' in the HSC SSP, mainly using the new $u$-band filter on MegaCam (Figure~\ref{fig1}), to a total area coverage of 18.6 deg$^2$.  Because of this large area coverage, CLAUDS is about 0.7 magnitude shallower than MUSUBI.  However, CLAUDS also included our $u^\ast$ data and $u^\ast$ data in the CHFT archive acquired by various teams previously, and reached the same depth as MUSIBI in the COSMOS and SXDS fields (aka. CLAUDS ``UltraDeep'' fields).  Here we provide our independent reduction and calibration of  the CFHT MegaCam images, and release the images and reference catalogs to the community\footnote{\url{http://www.asiaa.sinica.edu.tw/\~whwang/musubi}}.  The foci of this paper will be to describe the dataset in details and to compare our source catalogs with other publicly available catalogs in COSMOS and SXDS.

In Section~2, we provide an overview of the MUSIMI observations. The data reductions, calibrations, and resulting data qualities are described in Sections 3, 4, and 5, respectively. In Section 6, we provide details about the released data products. In Section 7, we showcase examples of sciences, specifically the photo-$z$ and the evolution of green-valley galaxies, enabled from the combined MUSUBI and HSC datasets. Throughout the paper, we adopt the AB magnitude system \citep{oke83}, and the standard $\Lambda$CDM cosmological parameters of $H_0=70$ km s$^{-1}$ Mpc$^{-1}$, $\Omega_m = 0.3$, and $\Omega_\Lambda=0.7$.

\section{Observations}
Our MUSUBI team based in Taiwan carried out extremely deep $u^\ast$-band imaging of the COSMOS and the SXDS fields with MegaCam on CFHT, during semesters from 2012B to 2016B (PI: S.\ Foucaud and W.-H.\ Wang).  All observations were conducted under the queue observing mode, with active atmospheric extinction and seeing monitoring. The queue mode allows us to only observe when the $r$-band seeing is better than $0\farcs8$, corresponding to average to good seeing on Maunakea.  However, the $u^\ast$-band seeing is typically about $0\farcs14$ worse than that at $r$-band, so our actual image quality is typically worse than $0\farcs8$.  We requested the queue system to only observe our targets under photometric conditions.  The CFHT queue service observing team also acquired standard star observations and twilight flats when the conditions were suitable.

Our primary goal is to support the two Ultra-Deep fields in the HSC SSP. We therefore only imaged the 1-deg area of the center of each of the COSMOS and SXDS fields with one MegaCam pointing, to reach the highest sensitivity given the time constraint.  Our exposures were dithered to cover the gaps between the MegaCam CCDs.  From semester to semester, we slightly changed the dither pattern and the pointing center for each field, to further even out the sensitivity distribution and to slightly expand the maps.  Typical exposure times adopted in our observations are from 480 to 720 seconds, to minimize the readout and dither overheads, and to take advantage of the dark sky.  Only a small fraction of the exposures are shorter, from 240 to 320 seconds, to accommodate observing conditions that are less ideal.  Over the course of 3.5 years, we had accumulated 20.3 hr of exposures for the COSMOS field, and 41.8 hr for the SXDS field.  

In addition to our own observations, in our data reduction, we also include all available archival data that cover our field centers.  The exposure times of the archived data are from 120 to 660 seconds.  The seeing and extinction variations in the archived data are also larger than ours.  Among all our data and the archived data, 5.2\% of exposures for COSMOS were taken under thin cirrus where the extinction exceeds 0.1 magnitude, while the conditions for SXDS exposures remain all photometric. To minimize the impact of poorer data from the archive, our data reduction (Section~\ref{sec_reduction}) down-weighted data taken under poorer conditions automatically. Moreover, we visually inspected all exposures to remove obviously bad ones.  The inclusion of the archival data brings the total amount of data to 78.1 hr for COSMOS and 60.5 hr for SXDS.  However, the archival data were taken by various teams with various imaging strategy.  A substantial fraction of the past observations were to mosaic much wider areas, rather than single focused pointings.  Therefore, in our final reduced maps, only the map centers receive more than 60 hr of exposure times for both fields.  

\begin{figure}
\epsscale{0.6}
\plotone{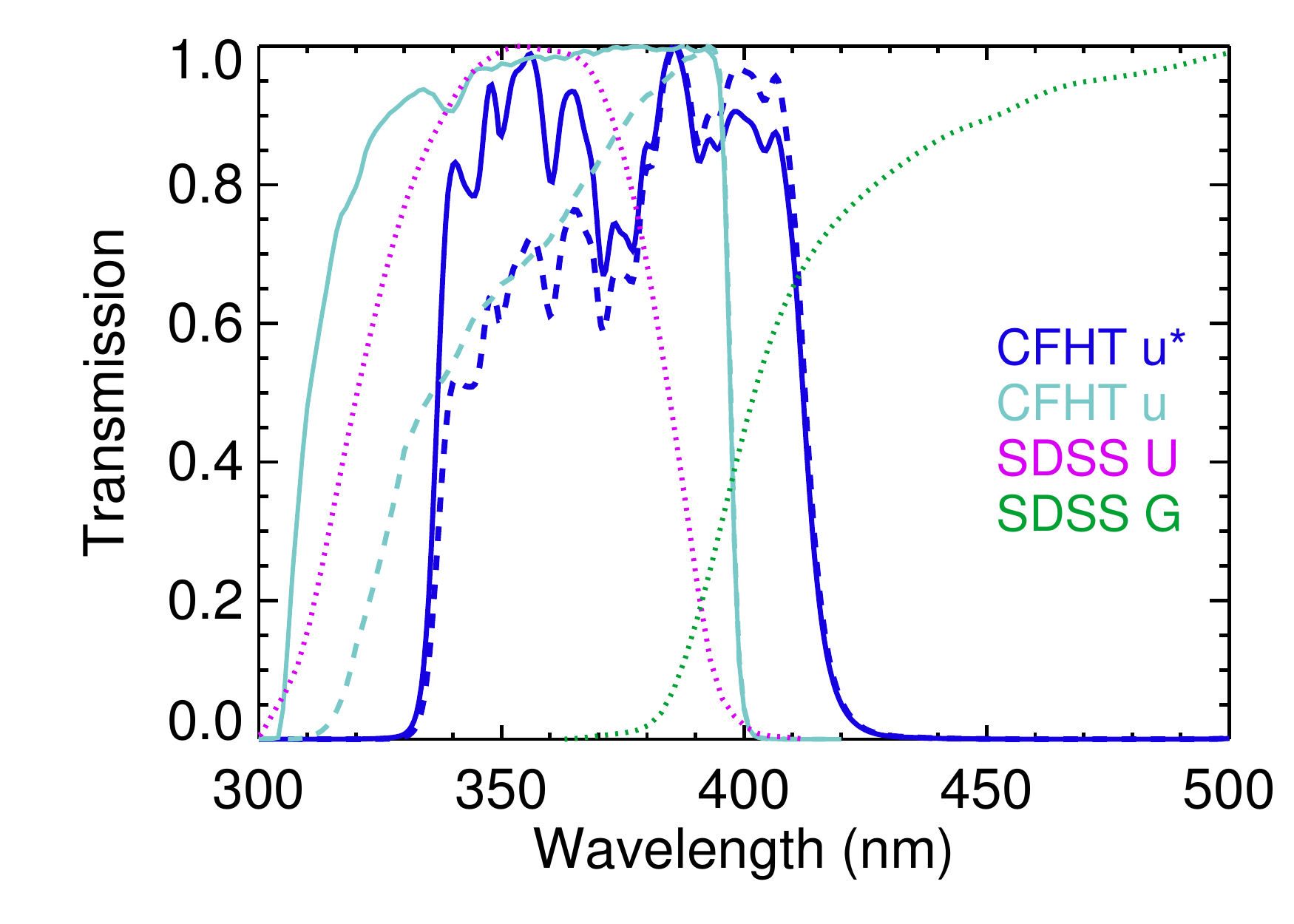}
\caption{Transmission profile of the CFHT MegaCam $u^\ast$ filter used in MUSUBI. The solid blue curve is the profile of the filter, while the dashed blue curve is the profile combined with the telescope/camera transmission and CCD quantum efficiency.  For readers' reference, we also show the profiles for the new CFHT $u$ filter (solid cyan curve), the new CFHT $u$ filter combined with telescope/camera transmission and CCD quantum efficiency (dashed cyan curve), the SDSS $U$ filter (dotted magenta curve), and the SDSS $G$ filter (dotted green curve). All profiles shown here do not contain the effect of atmospheric absorption.
\label{fig1}} 
\end{figure}

All our and archived imaging was conducted using the old $u^\ast$ broad-band filter.\footnote{Since semester 2015A, a new set of broad-band filters were introduced to MegaCam. Its filter transmission curves are shown in Figure~\ref{fig1}. In CFHT's documentation, the old $u^\ast$-band filter is referred as $u^\ast$ or $uS$, while the new $u$-band filter is bluer. Here we simply refer our filter as $u^\ast$ and the readers should not be confused this with the new $u$ filter.} Its filter transmission curve is presented in Figure~\ref{fig1}.  The filter itself has a central wavelength of 375 nm, and a bandwidth of 74 nm at 50\% transmission.  The average wavelength slightly shifts to 379 nm after taking the telescope/camera optics transmission and CCD quantum efficiency into account.  Atmospheric transmission would further shift the average wavelength to longer wavelengths and this is airmass-dependent.  Figure~\ref{fig1} also compares our filter with the two adjacent SDSS filters, denoted as $U$ and $G$. Because our $u^\ast$ filter is substantially different from $U$, we cannot directly use the SDSS data to calibrate our $u^\ast$-band photometry (Section~\ref{sec_photo_cal}).

\section{Data Reduction}\label{sec_reduction}
All the MegaCam data were preprocessed by the CFHT Elixir system\footnote{\url{http://www.cfht.hawaii.edu/Instruments/Imaging/MegaPrime/}} \citep{magnier04} to remove instrumental features. This includes overscan and bias subtraction, flat fielding to correct for pixel-to-pixel and CCD chip-to-chip  sensitivity/illumination non-uniformity, and sky subtraction.  The file headers also contain updated astrometry and photometric zero points derived by the Elixir system.  Our data reduction starts from the Elixir preprocessed files.

We used subroutines in the SIMPLE Imaging and Mosaicking PipeLinE \citep[SIMPLE,][]{wang10,wang10b} to further process and mosaic the Elixir preprocessed CCD images.  We divided images into groups. Images taken by the same team (i.e., similar dithering and exposure), within the same semester, and from the same CCD were grouped and processed together.  We first conducted additional passes of sky background subtraction, to remove residual image background, which often share a common pattern among the grouped exposures.  In the first pass, we masked detected objects in each individual exposure, derived a median image from the group, and subtracted it from each exposure.  Then on each individual exposure, we masked detected objects again, fitted the masked image with a third-degree 2D polynomial surface, and subtracted fitted surface from the image. The object masking was done by first smoothing the image with a $3\times3$ square tophat kernel and then masking pixels that exceed 3 $\sigma$ locally, where $\sigma$ is measured in the unsmoothed image. The above procedure almost always leads to sufficiently flat sky in the images, such that the stacking and mosacking do not produce sharp background offsets along the CCD boundaries. The only exception is the rare cases where extremely bright stars prevent good polynomial background models or median sky models.

The grouped and sky-subtracted images were then fed to SExtractor \citep{bertin96} to generate a source catalog for each of them.  The photometry of compact objects that have signal-to-noise raio (S/N) higher than 10 in the SExtractor catalogs were compared, so images taken under non-photometric conditions can be identified. These images were re-normalized so their photometry matches the others.  The atmospheric absorption derived for these images were also used to weight these images in the image stacking step according to the atmospheric extinction.  The coordinates of objects in the SExtractor catalogs were compared among the images, to derive the exact amounts of the dither offsets and the optical distortion.  The optical distortion can be derived because the displacement of stars among the dithered images as a function of position in the images is the first-order derivative of the distortion function (e.g., \citealp{anderson03}; see more details in \citealp{wang10}).  The distortion correction was then applied to each image to project them to a tangential sky.  All our images in the same field share the same projection center, position angle (0), and image scale ($0\farcs186$, the native pixel scale of MegaCam). The distortion correction and the sky projection take into account the change in area for each pixel in a way that source fluxes are conserved. This is required to achieve a uniform absolution flux calibration across the entire field (``photometric flat-fielding''). We will discuss the absolute flux calibration in Section~\ref{sec_photo_cal}.

Two passes of cosmic ray removal were applied to the images.  First, on each exposure, a bright pixel was masked if it exceeds 4 $\sigma$ in its  $9\times9$-pixel neighborhood.  These criteria were carefully chosen such that only spikes much narrower than the point-spread function (PSF) would be masked and stars would not be affected. Second, in a group of dithered, distortion-corrected, and sky-projected images, pixels that share the same sky coordinates were compared against each other.  Outliers in the pixel brightness distribution were considered as spurious and masked.  This is repeated for every independent pixel on the sky.  Typically several tens of dithered images are in a group, meaning that tens of pixels were compared against each other.  Only around the boundary of the CCD where the number of overlapping pixels decreases, this method becomes less effective.

\begin{figure*}[t!]
\epsscale{1.15}
\plotone{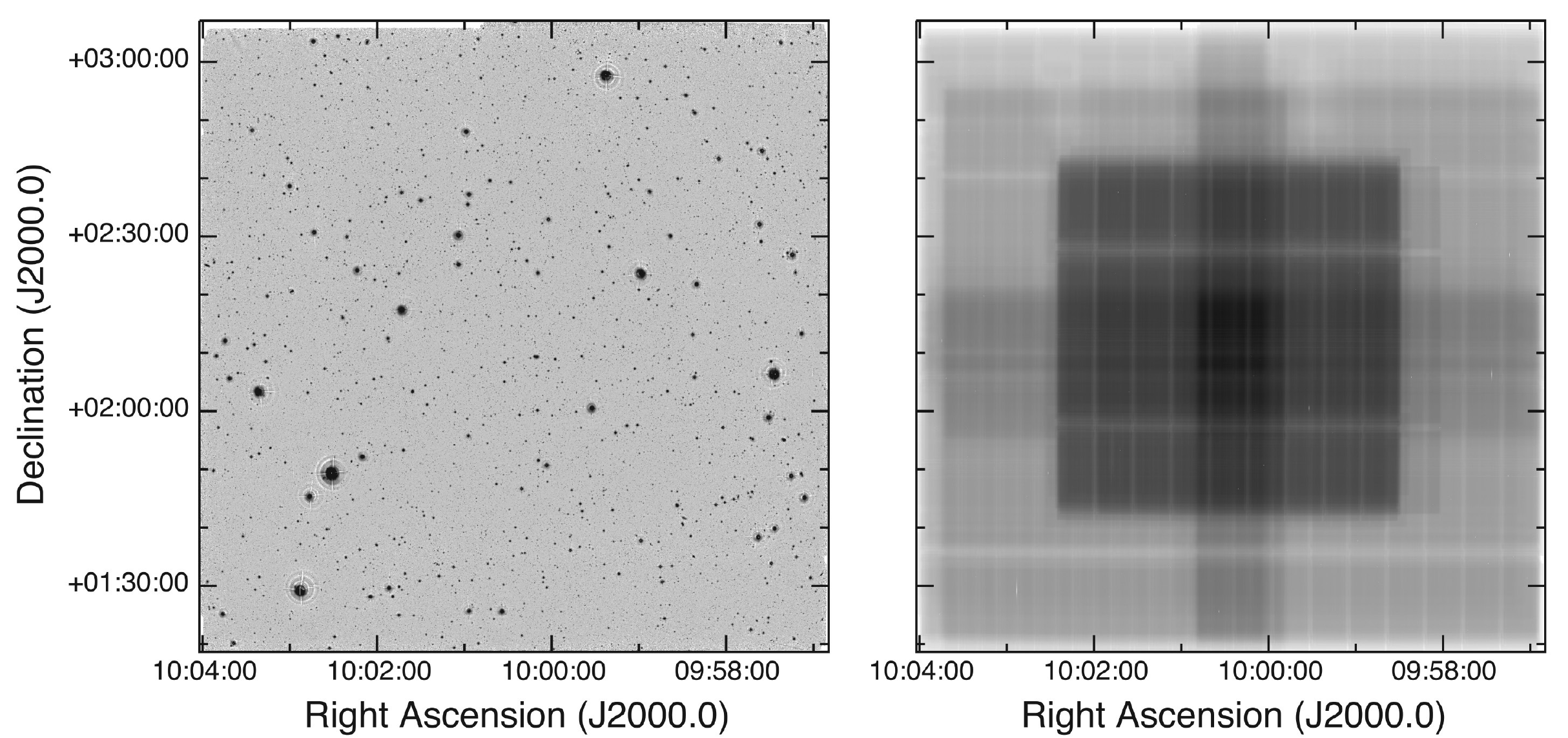}
\caption{Image of the COSMOS field (left) and its weighted exposure distribution (right). The brightness scales are inverted. The entire area shown here has a size of $108\arcmin \times 108\arcmin$. The image shows that this field is relatively free from bright stars and the background subtraction in our reduction is excellent. The exposure map shows the very different mosaicking strategy adopted by various teams, producing an extremely deep core of approximately $11\arcmin \times 14\arcmin$ at the center and a deep area of approximately $60\arcmin \times 60\arcmin$.
\label{fig2}} 
\end{figure*}

\begin{figure*}[t!]
\epsscale{1.15}
\plotone{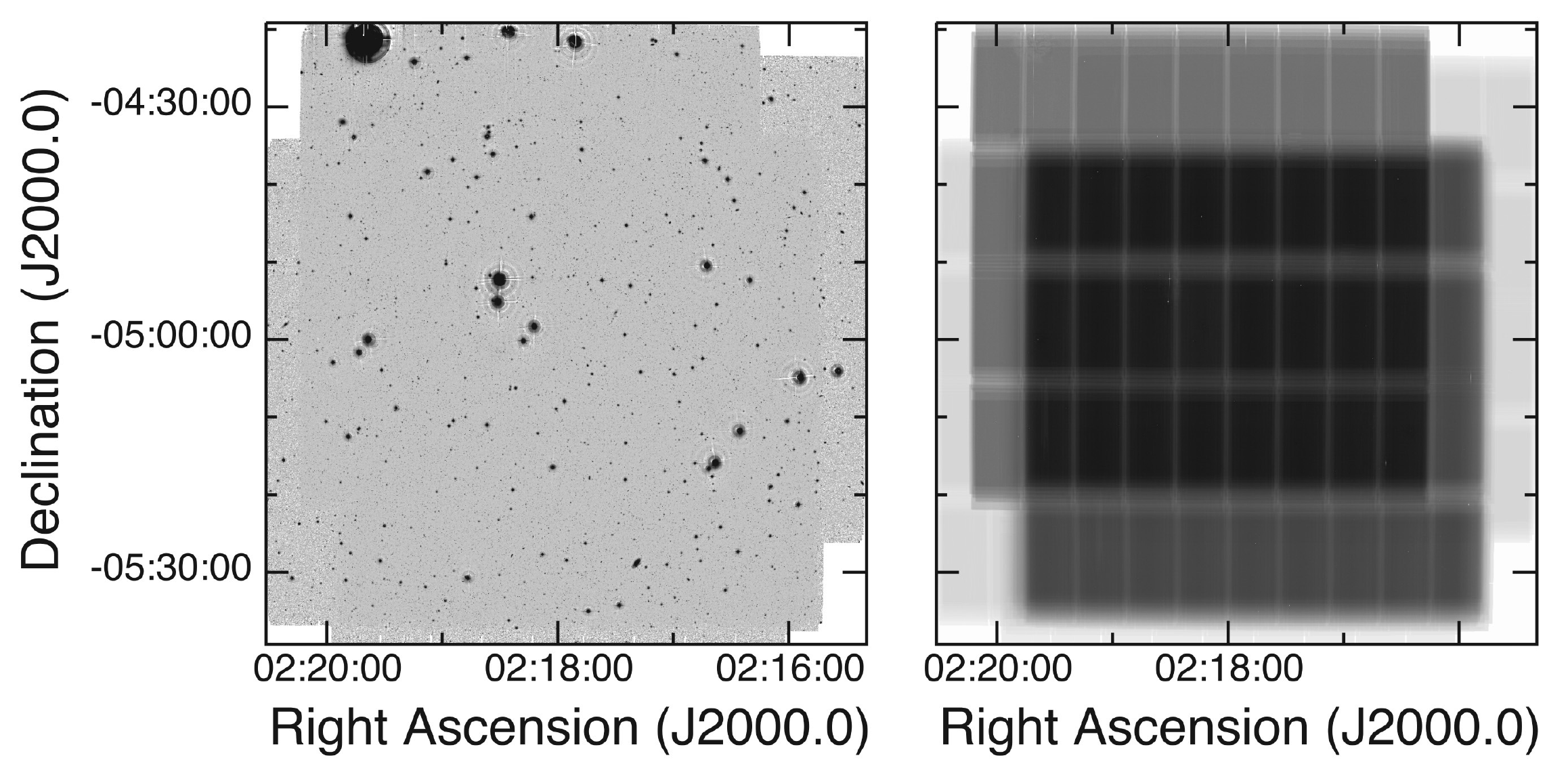}
\caption{Image of the SXDS field (left) and its weighted exposure distribution (right). The brightness scales are inverted. The entire area shown here has a size of $77\arcmin \times 80\arcmin$. The bright star near the north-western corner has $V=6.49$ and produces slightly less than ideal background subtraction in its neighborhood.  The exposure distribution is more uniform and the map size is smaller than the COSMOS field.  The deep area is approximately $52\arcmin \times 44\arcmin$.
\label{fig3}} 
\end{figure*}

After the above processing, the images were then average-stacked to form a deep image. In the stacking, the images were weighted according to their exposure time and atmospheric extinction. Then the stacked images from all MegaCam CCDs, projects, and semesters were mosaicked and stacked to from the final wide-field, ultra-deep images. In the final stacking and mosaicking, pixels that share the same sky coordinates were again compared against each other to remove residual cosmic ray hits that were not previously masked, before the images were combined. In Figure~\ref{fig2} and Figure~\ref{fig3}, we present the final mosaics for the COSMOS and SXDS fields, respectively, and their associated weighted exposure time maps.

\section{Calibration}\label{sec_calibration}

In this section, we outline how we achieved the calibrations for photometry and astrometry, and present the quality of the calibrations.

\subsection{Photometric Calibration and Quality}\label{sec_photo_cal}
\subsubsection{COSMOS}
Our general strategy is to tie our photometry to the well calibrated CFHT Supernova Legacy Survey (SNLS, e.g., \citealp{astier06,guy10}). The SNLS covered the COSMOS field with MegaCam and with the same $u^\ast$ filter used in MUSUBI.  \citet{betoule13} published improved photometric calibration of the SNLS, which took into account the varying passband of the filter across the field and photometric flat-fielding (also see \citealp{regnault09}).  The resultant photometric uniformity is 8 milli-magnitude at $u^\ast$, sufficient for the studies of high-redshift galaxies. We note that the Elixir data reduction mentioned in Section~\ref{sec_reduction} also adopts a photometric calibration matched to the SNLS calibration in \citet{betoule13}, since 2015. However, we include substantial amount of data taken prior to that, whose calibration was based on the SDSS calibration of \citet{smith02} transferred to the MegaCam system \citep{magnier04}.  To correct for any potential offset between the old MegaCam calibration and the new one, we applied the SNLS-based photometric calibration to each dataset individually before they were combined to form a deep image.  

To calibrate our COSMOS $u^\ast$ data, we adopt the ``uniform magnitude'' in the SNLS photometric catalog of \citet{betoule13}, which corrects for the varying filter passband across the field.  Then for our own data, we measured galaxy photometry using an aperture with $5\arcsec$ diameter. This is reasonably large for encapsulating the total fluxes of faint galaxies, as this is the aperture that gives fluxes closest to SExtractor auto-aperture fluxes on well detected compact objects for our PSF ($\sim0\farcs9$, see Section~\ref{sec_quality}).  This is also similar to the aperture adopted by the SNLS ($15\times$ Gaussian $\sigma$ of the PSF).  We calculated the photometric correction by comparing the two for data taken in each observing run.  The differences between our measured (Elixir-calibrated) photometry and the SNLS photometry in the various observing runs are always a few percents, even for data taken after 2015. This is likely caused by the difference in seeing and the aperture sizes adopted by Elixir. Our correction eliminates these offsets. The SNLS D2 pointing has an area of 1 deg$^2$, which only covers the central part of our 3 deg$^2$ map.  For the outer region that does not have SNLS photometry, we propagated the photometric solution that we obtained from the center using the overlapping regions between the central and outer pointings. Our calibrated images have a map unit of $\mu$Jy per pixel, which is equivalent to an AB magnitude zero point of 23.9.

\begin{figure}[t!]
\epsscale{0.6}
\plotone{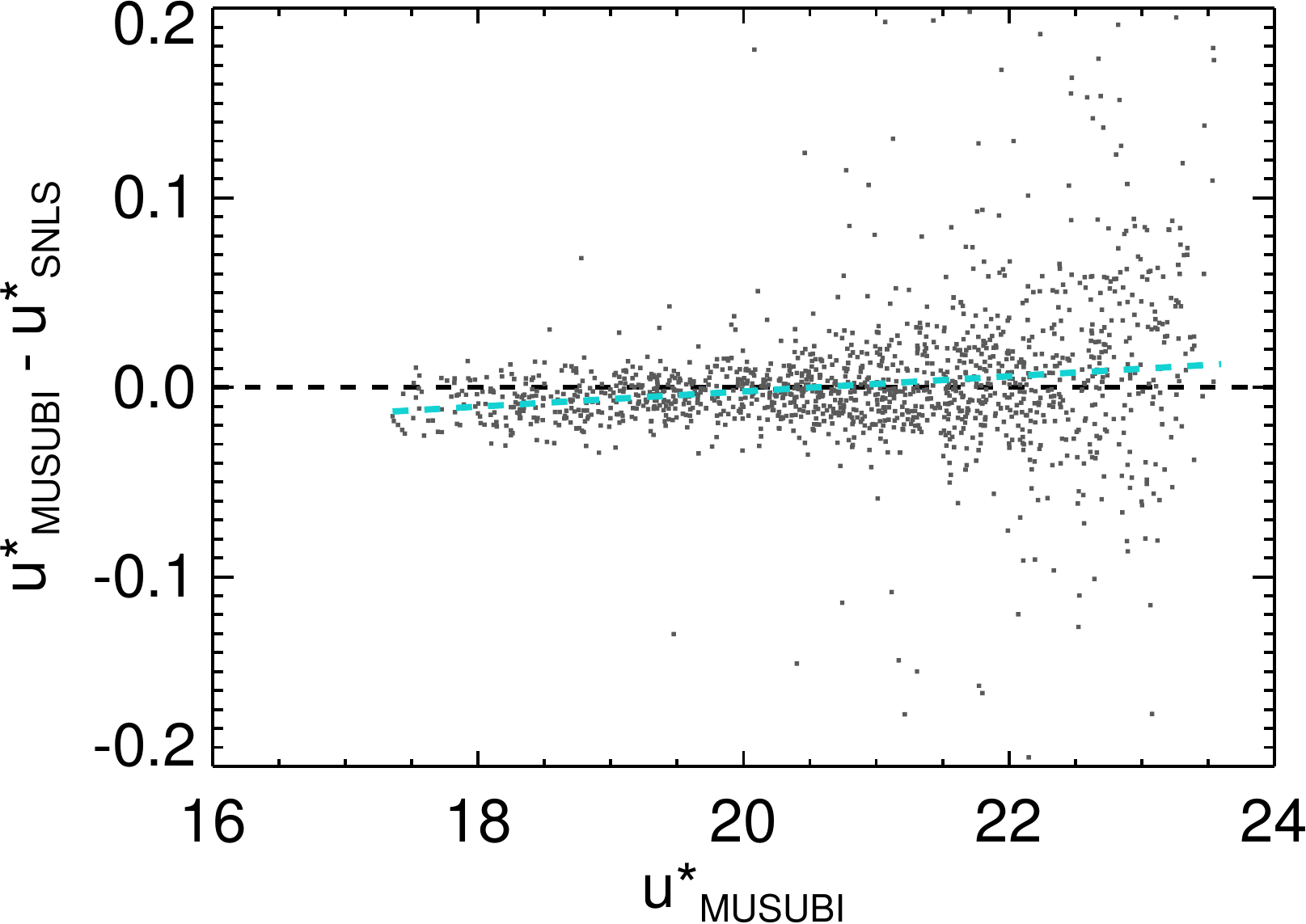}
\caption{Comparison between our COSMOS photometry and the SNLS photometry on objects with  $u^\ast$ errors less than 0.05 magnitudes in both catalogs.  We use $D=5\arcsec$ apertures from our catalog, which is comparable to the aperture size used in SNLS.  The median for the full sample is $-0.0018$.  The cyan dashed line is a linear fit to data within $y=\pm0.1$, suggesting that there is a tilt. The nature of this tilt is unclear, and it is nevertheless very small.
\label{fig_cosmos_cal}} 
\end{figure}

We demonstrate the calibration quality in Figure~\ref{fig_cosmos_cal}, in which we compare the photometry from our SExtractor catalog derived from our final mosaic (Section~\ref{sex_catalog}) with that from SNLS.  Overall, the two catalogs agree with each other, and the median offset is $\Delta u^\ast=-0.0018$. However, there seems to be a small tilt in the sequence in Figure~\ref{fig_cosmos_cal}, ndicated with the cyan dashed line.  The median offsets are $-0.0039$ at $u^\ast<21$ and $0.0040$ at $u^\ast>21$.  There is not an obvious explanation to this tilt, but the $\pm0.4\%$ offset should not have practical impacts to most science cases. The largest uncertainty of calibration should come from the fact that there are typically only 20 to 30 SNLS objects available for calibrating each MegaCam CCD. Using the dispersion in Figure~\ref{fig_cosmos_cal}, which is 0.055 magnitude, we estimate that the calibration uncertainty would be approximately 0.01 magnitude, which is still quite good.  The results here show that we have reached excellent calibration relative to the SNLS photometry in the COSMOS field.

\subsubsection{SXDS}
In the SXDS field, our pointing and the SNLS D1 pointing are separated by about $2\arcdeg$. Therefore, unlike the COSMOS field, we cannot directly use the SNLS photometric catalog of \citet{betoule13} to calibrate our SXDS data.  Here we rely on the SDSS $U$ and $G$ photometry in our field, but converted to the SNLS $u^\ast$ photometric system.  

\begin{figure}[t!]
\epsscale{0.6}
\plotone{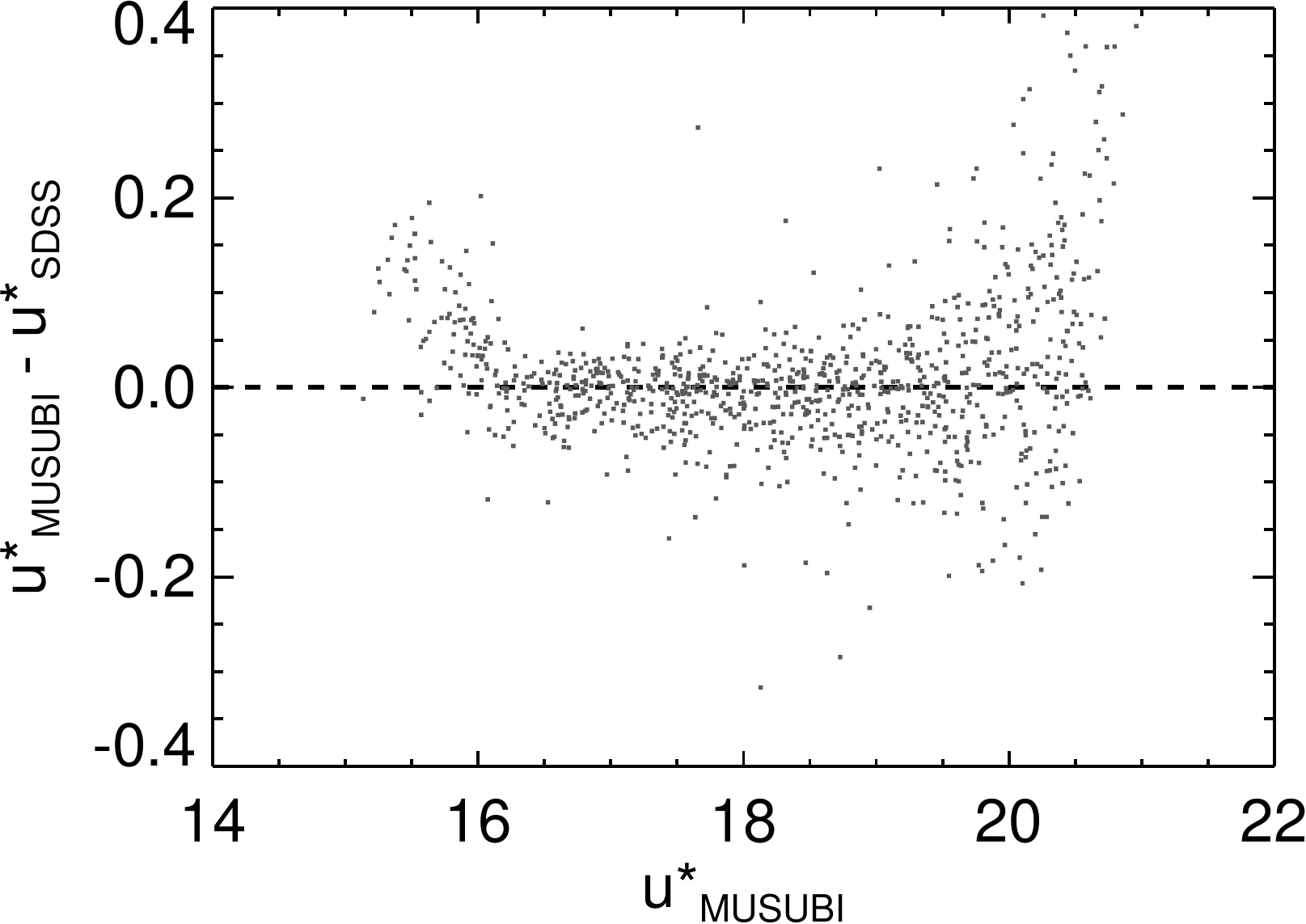}
\caption{Comparison between our SXDS photometry and the synthetic SDSS $u^\ast$ photometry on objects with SDSS $U<21$.  We use $D=5\arcsec$ apertures from our catalog.  The median at $u^\ast>16.5$ is 0.012.  At $u^\ast<16.5$, nonlinear effects start to show up in the MegaCam data.   Such bright objects were not used for calibration. They are shown here only to show the nonlinear effects.  
\label{fig_sxds_cal}} 
\end{figure}

First, we selected blue stars and galaxies from the SDSS DR12 catalog using their $U-G$ colors. This avoids passive galaxies and late-type stars, whose strong 4000~\AA\, breaks can induce large color terms in the $U$ and $G$ bands, particularly for galaxies, whose color terms can be redshift-dependent.  We only use galaxies brighter than $U=21$ to avoid large photometric uncertainties.  By cross-matching the selected SDSS galaxies and SNLS galaxies, we derived the following conversions with least-square fitting:
\begin{equation}
u^\ast = U - 0.187  (U-G) - 0.1443.
\label{eq_sxds_conversion}
\end{equation}
To better understand this relation, we further picked a sub-sample of galaxies with SDSS spectroscopy and conducted SED fitting to their SDSS photometry at their redshifts.  We then integrated their fitted spectra using the $u^\ast$ filter profile to derive their $u^\ast$-band magnitudes. The result is consistent with the above empirical relation, but noisier because of the uncertainties associated with the SED fitting and the smaller sample size. We therefore conclude that Eq.~\ref{eq_sxds_conversion} correctly describes blue stars and galaxies within the SDSS detection limit. We used this relation to derive synthetic $u^\ast$ photometry of SDSS galaxies in our SXDS field to calibrate our CFHT data.  In Figure~\ref{fig_sxds_cal}, we show a comparison between our SExtractor $u^\ast$ photometry and the synthetic SDSS $u^\ast$ photometry. Overall, the agreement between the two is very good, though slightly worse than the case for COSMOS based directly on SNLS.The median offset at $u^\ast>16.5$ is 0.012 magnitude, while the dispersion is 0.10.  This leads to a calibration uncertainty of roughly 0.02 magnitude for each MegaCam CCD.  This is about $2\times$ as worse than the case for COSMOS.

\subsection{Astrometric Calibration and Quality}

We tie our astrometric system to that of the Gaia \citep{gaia16} data.  During our our data reduction, the individual exposures from each MegaCam CCD were corrected for distortion, tangentially reprojected on to a common sky plane for each field, and then stacked/mosaicked with other exposures.  The reprojection aligned our detected objects with the coordinates in the Gaia DR2 catalog \citep{gaia18,lindergren18}.  In this process, we only used Gaia objects that are brighter than Gaia $G=20.5$ and have proper motions measured to be less than 30 milli-arcsec (mas) year$^{-1}$.  The projection center for COSMOS is R.A.~(J2000.0) = 09:59:59.59, Decl.~(J2000.0) = +02:12:08.18. The projection center for SXDS is R.A.~(J2000.0) = 02:18:00.00, Decl.~(J2000.0) = $-$05:00:00.00.  These positions were roughly determined from the common center of the various pointings of our observations and the archival data.  The projected images have position angles of 0.0 on the sky and maintain the native MegaCam pixel size of $0\farcs186$ at the projection centers.

\begin{figure*}[t!]
\epsscale{1.1}
\plotone{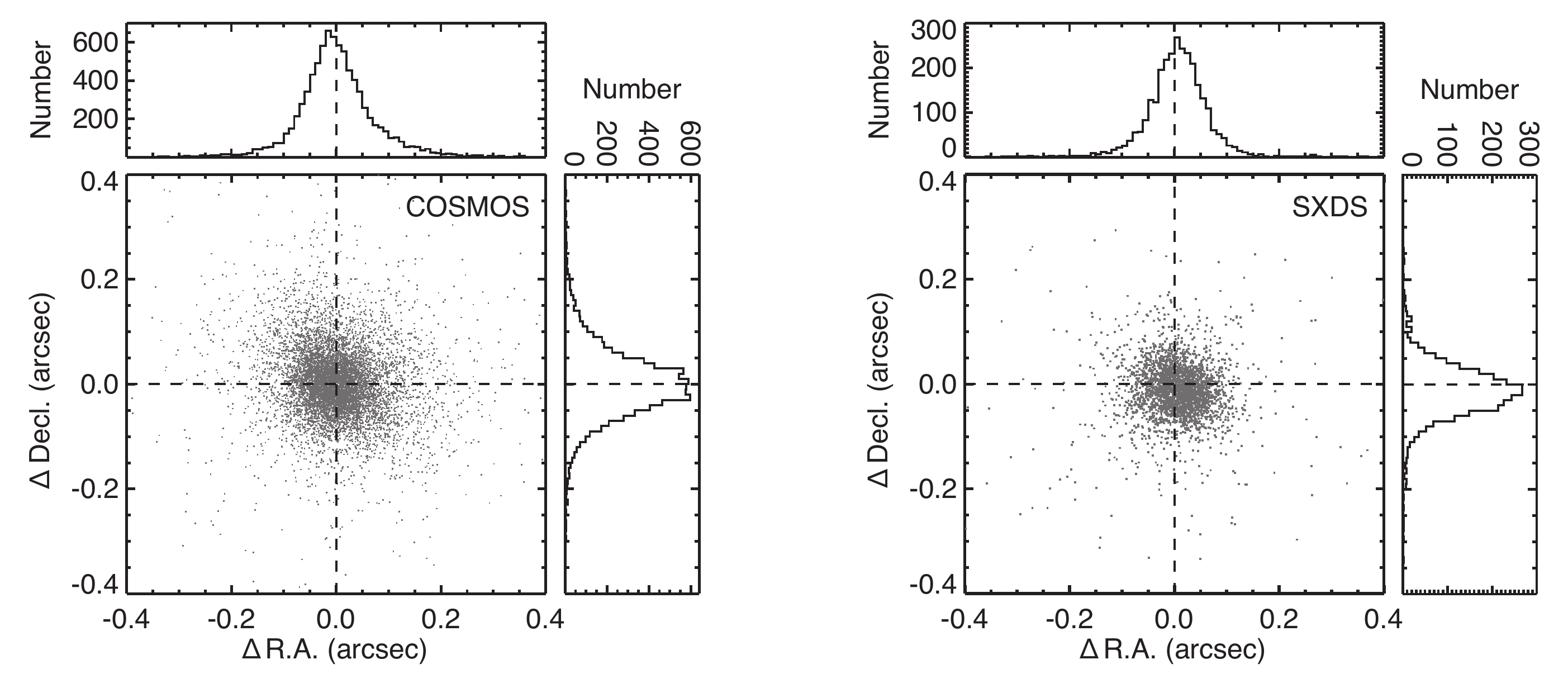}
\caption{Astrometry of our data relative to the Gaia DR2 catalog. The positional offsets shown here are MUSIBI $-$ Gaia.  The mean offsets are around 10 mas in both fields.  The dispersions are approximately 70 mas for COSMOS and 60 mas for SXDS.
\label{fig_astrometry}} 
\end{figure*}

To examine the quality of the astrometry calibration, we show comparisons of the source positions in our final catalog against their Gaia positions in Figure~\ref{fig_astrometry}.  For the COSMOS field, where 8360 sources are included in the comparison, the mean positional offsets along R.A.\ and Decl.\ are both at mas levels, practically consistent with zero.  The dispersions in the offsets are 67 mas for R.A.\ and 74 mas for Decl., both acceptably small.  For the SXDS field, where 2896 sources are included in the comparison, the mean offsets along the R.A.\ and Decl.\ are 6 mas and $-11$ mas, respectively.  Therefore, there is a slight tendency of 10 mas for our positions to offset to the south-east relative to Gaia.  The dispersions in the offsets for SXDS are 52 mas for R.A.\ and 61 mas for Decl., about 20\% smaller than those in the COSMOS field.  This 20\% difference is likely a consequence of the smaller SXDS coverage, so the required distortion correction including reprojection is smaller.  If we only use the brightest 15\% of unsaturated sources ($u^\ast=17$--19), the dispersions are reduced by about 14\% (COSMOS) and 5\% (SXDS).  These relatively small improvements suggest that the positional errors relative to Gaia are not S/N-driven.  This could be the fundamental limit of our overall methodology and pipeline capability. We conclude that the systematic errors in our astrometric calibration is nearly negligible, while the uncertainties in the source positions measured from our images are at a small 50--70 mas level for bright sources that are not limited by S/N.

\section{Data Quality}\label{sec_quality}

\subsection{Sensitivity}

To examine our imaging sensitivity, we selected sources detected at $5\pm0.1~\sigma$ with SExtractor auto-apertures, and calculated their median value. The resultant median 5 $\sigma$ limiting magnitudes are 27.19 for the COSMOS and 27.68 for the SXDS.  

\begin{figure*}[t!]
\epsscale{1.1}
\plotone{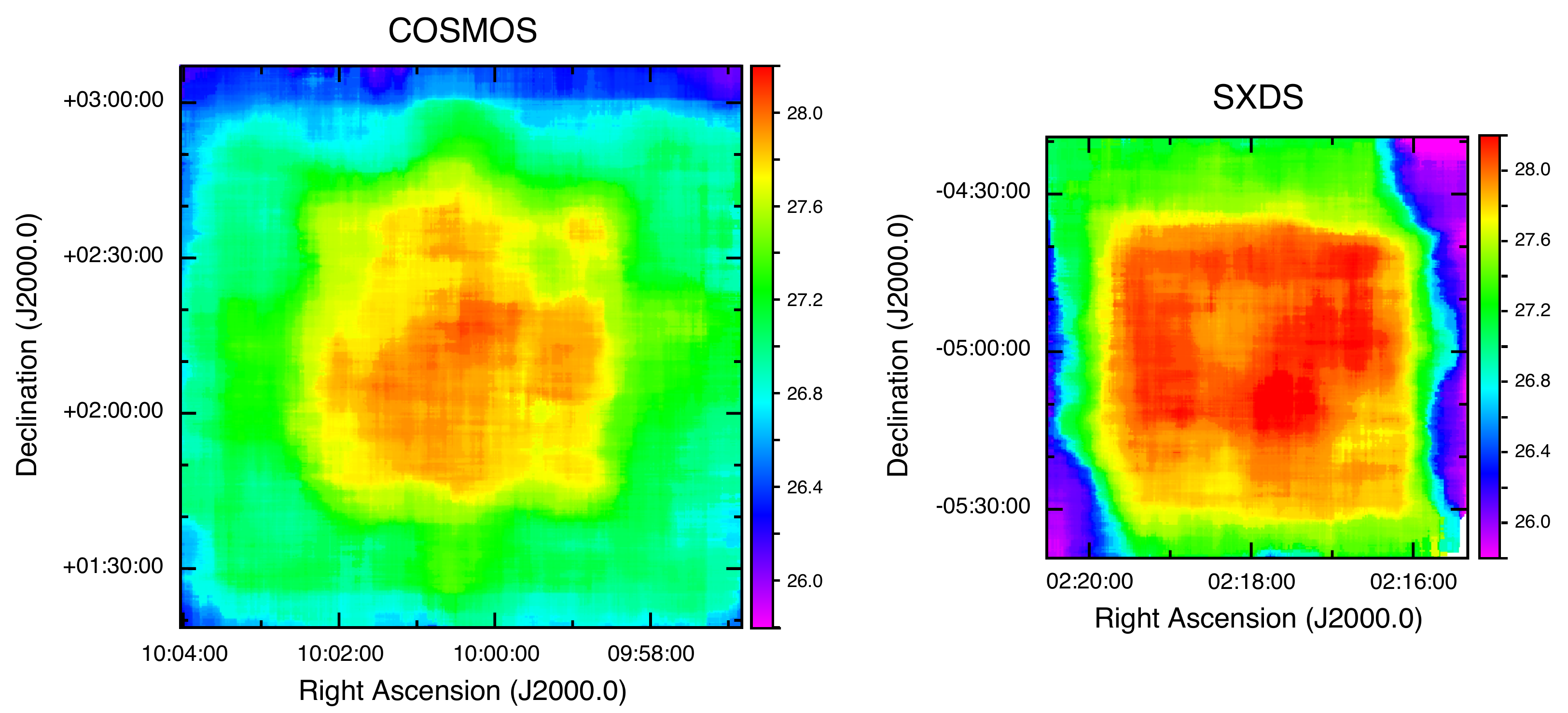}
\caption{$u^\ast$-band sensitivity distribution of our survey. The color shows the median $u^\ast$ magnitudes of all sources detected at 5 $\sigma$ with SExtractor auto-apertures. The angular scales for the two fields are identical.   It can be seen that the deep regions (yellow to red) in both fields are slightly less than 1 deg$^2$, and reach 27.7 in the COSMOS and 27.8 in the SXDS.  For point-like sources, the limiting magnitudes are roughly 0.3 mag deeper.
\label{fig_lim_dist}} 
\end{figure*}

The values quoted above are for the whole fields. However, as shown in Figure~\ref{fig2} and Figure~\ref{fig3}, our integration time distributions are highly non-uniform, because of the different imaging strategies adopted by the various teams.  Therefore, the sensitivity distributions are also highly non-uniform.  We measured the 5-$\sigma$ limiting magnitudes in small areas. The results are shown in Figure~\ref{fig_lim_dist}. In the deepest regions in the COSMOS and SXDS, we reach $u^\ast=28.1$ and $u^\ast=28.4$, respectively.  In the $\sim1$ deg$^2$ relatively deep regions (yellow to red colors in Figure~\ref{fig_lim_dist}) in the COSMOS and SXDS, we reach 27.7 and 27.8, respectively. These two  $\sim1$ deg$^2$ regions were referred as ``CLAUDS UltraDeep'' regions by the CLAUDS team. These are by far the deepest 1 deg$^2$ fields for $u^\ast$ and similar $U$ bands.

All the above-mentioned limiting magnitudes were derived from all detected sources, among which many are extended objects. If we only select point-like objects whose measured sizes are $<1\farcs2$, the limiting magnitudes would become deeper by  $\gtrsim0.3$.  Also, if we fixed the aperture diameters to be $2\arcsec$, which favors compact objects, the limiting magnitudes would become 0.24 and 0.45 magnitude deeper in the COSMOS and SXDS, respectively. The difference between the two fields when switching to $2\arcsec$ is caused by the different image quality (next subsection).

\subsection{Image Quality}

\begin{figure}[ht!]
\epsscale{0.6}
\plotone{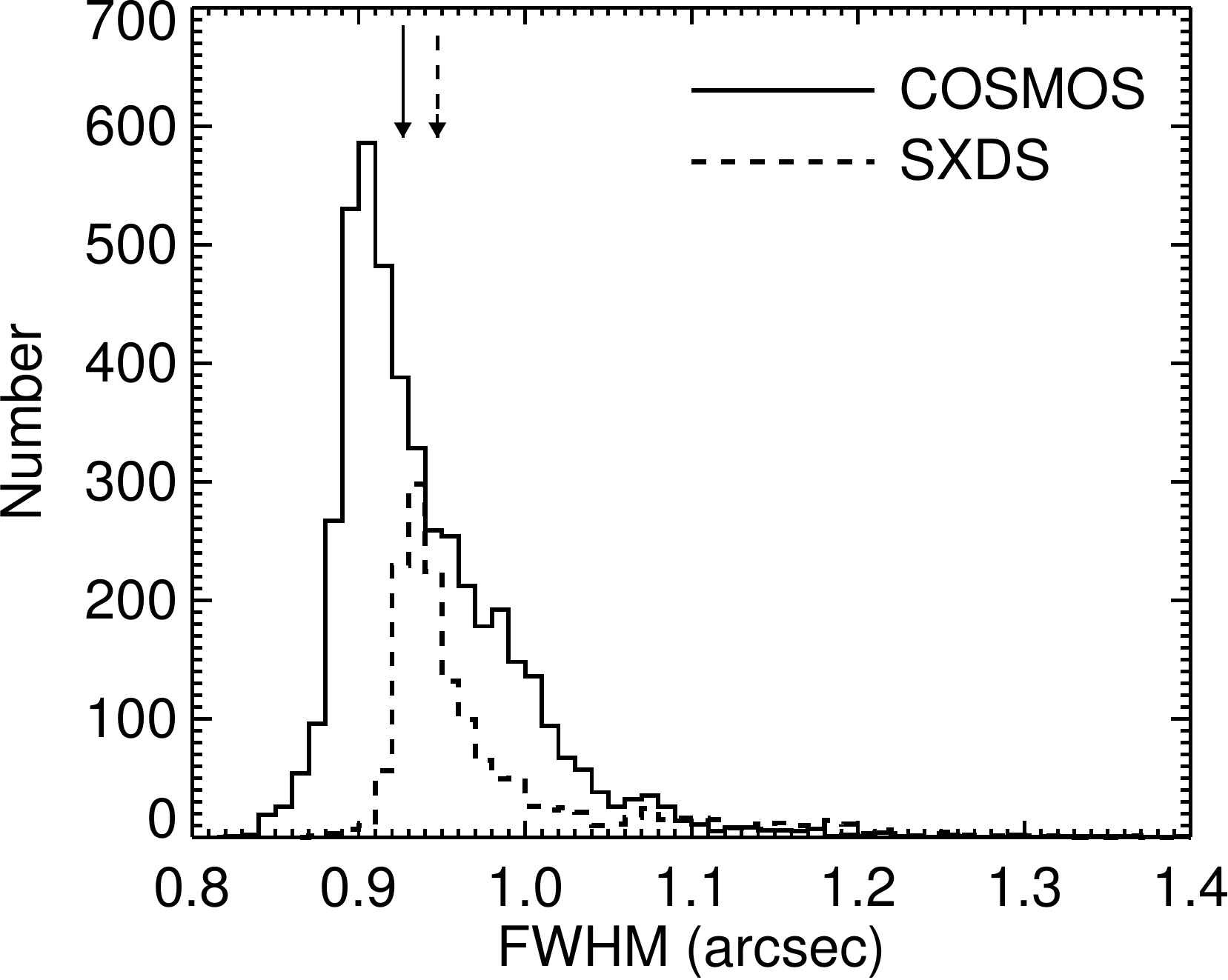}
\caption{$u^\ast$-band FWHM measured on SDSS stars and quasars with $u^\ast=17$--22 in the 
COSMOS (solid histogram) and SXDS (dashed histogram) fields. The two downward arrows indicate the medians of the distributions.
\label{fig_fwhm}} 
\end{figure}

To evaluate the image quality, we selected SDSS photometric stars and spectroscopic quasars in the COSMOS and SXDS fields
with $u^\ast>17$ and measured their FWHM with SExtractor.  We only selected $u^\ast>17$ objects to avoid nonlinear effects
in the MegaCam data. The cutoff of SDSS detection limits is roughly $u^\ast=22$.  The distributions of the FWHM values are shown in Figure~\ref{fig_fwhm}.  The median values are $0\farcs926$ for the COSMOS and $0\farcs947$ for the SXDS, indicated by the two arrows in Figure~\ref{fig_fwhm}.  

\begin{figure*}[ht!]
\epsscale{1.1}
\plotone{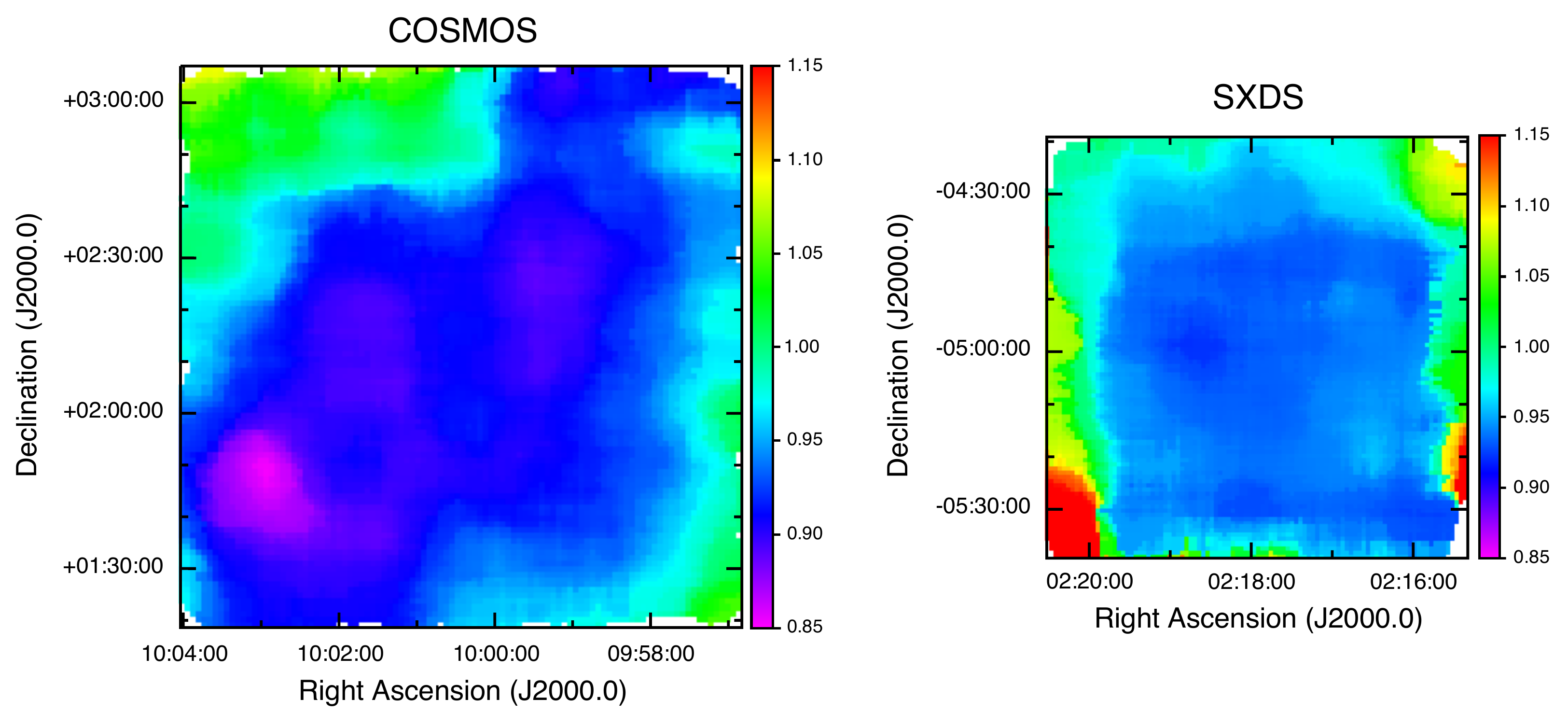}
\caption{Distribution image quality of our survey. The color shows the median FWHM (in arcsec) of SDSS stars or quasars with $u^\ast=17$--22. The angular scales for the two fields are identical.  In the 1 deg$^2$ deep regions shown in Figure~\ref{fig_lim_dist}, the typical image quality is $\sim0\farcs93$ for both fields. In some regions in COSMOS, especially the south-eastern corner, the image quality is significantly better. 
\label{fig_fwhm_dist}} 
\end{figure*}

Like the situation for sensitivity, the image quality in the two fields is not uniform.  This is reflected in the asymmetric histograms in Figure~\ref{fig_fwhm}.  The histogram for FWHM becomes more symmetric and sharply peaked if we only look at small regions in both fields.  We therefore measured the medians of the FWHM distributions in small areas, and show the results in Figure~\ref{fig_fwhm_dist}. Overall, the center of the fields that are covered by our own deep imaging have better image quality, while the outer parts covered by various previous teams have larger image quality variation. The central 1 deg$^2$ region in the COMOS field has FWHM of $0\farcs88$--$0\farcs92$, while the central 1 deg$^2$ region in the SXDS has FWHM of 
$0\farcs92$--$0\farcs95$.

\begin{deluxetable}{lr}
\tablewidth{0pt}
\tablecaption{SExtractor Parameters \label{sexpara}}
\tablehead{\colhead{Parameter} & \colhead{Value}}
\startdata
DETECT\_MINAREA 	& 4 \\
DETECT\_THRESH    	& 1.0 \\
ANALYSIS\_THRESH  	& 1.2 \\
FILTER           			& Y \\
FILTER\_NAME		& gauss\_2.5\_5x5.conv \\
DEBLEND\_NTHRESH  	& 64 \\
DEBLEND\_MINCONT  	& 0.000001 \\
CLEAN            			& Y \\
CLEAN\_PARAM      		& 0.1 \\
SEEING\_FWHM      	& 0.93 \\
PIXEL\_SCALE			& 0.186 \\
MAG\_ZEROPOINT		& 23.9 \\
PHOT\_APERTURE		& 5.38, 8.06, 10.75, 16.13, 21.51, 26.88\\
BACK\_SIZE        		& 32 \\
BACK\_FILTERSIZE  	& 6 \\
BACK\_TYPE        		& AUTO \\
BACKPHOTO\_TYPE   	& LOCAL \\
BACKPHOTO\_THICK  	& 32 \\
WEIGHT\_TYPE      		& MAP\_WEIGHT \\
WEIGHT\_THRESH    	& 1000 \\
\enddata
\end{deluxetable}

\section{Source Catalogs}\label{sex_catalog}

In our data release, we provide reference catalogs along with the images. These catalogs can be readily used for scientific studies. However, the users may need to generate their own catalogs if there are special requirements on the photometry, completeness, or sample purity, or if there are needs of photometry based on position priors.

\subsection{Catalog Generation}
\begin{figure*}[ht!]
\epsscale{1.1}
\plotone{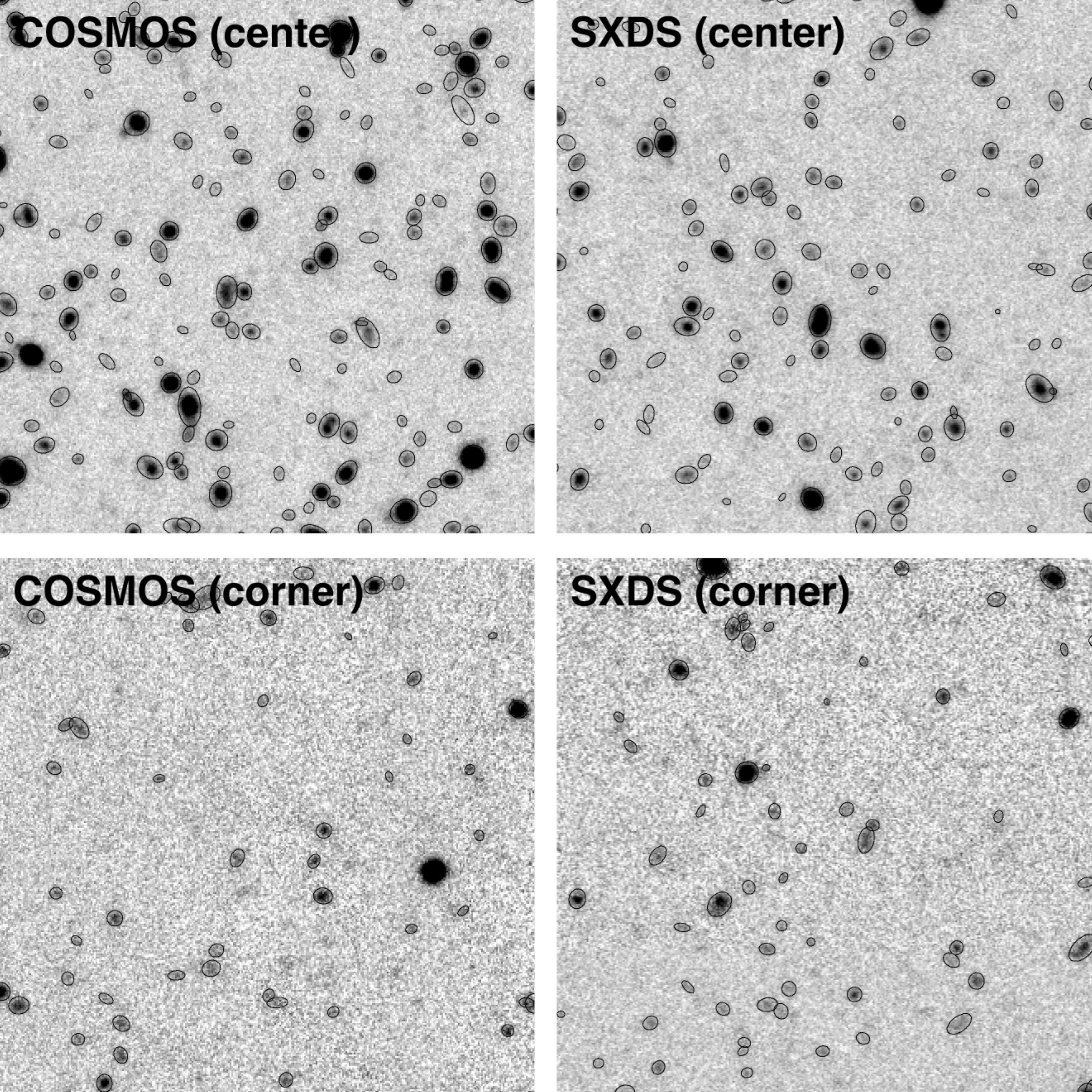}
\caption{Illustrations of source extraction in our reference catalogs, to give readers a general idea about our detection aggressiveness. The center and a shallow corner of each field are shown. Each panel is $1\arcmin$ in size. The two central panels have identical brightness scales, while the two corner panels have $3\times$ larger brightness scales because of the higher noise. The ellipses are drawn according to SExtractor's shape parameters ($A$, $B$, and $THETA$) for objects with S/N $> 5$ (auto aperture) in our reference catalogs.  The major and minor axes of the ellipses are made $2\times$ larger so faint objects behind them can be more easily seen. With careful examination, one may notice about 20 (center panels) or 10 (corner panels) objects in each panel missed by SExtractor, either because of their close proximity to birghter objects, or because of their extreme faintness. One may also notice less than a handful of ellipses in each panel that are more likely noise spikes rather than convincing detections. Our source extraction parameters are chosen to reach a balance between detection completeness and spurious detections.
\label{fig_extraction}} 
\end{figure*}

We used SExtractor ver.2.5.0 to generate the reference catalogs.  The key SExtractor parameters are listed in Table~\ref{sexpara}. Because the two fields have very similar properties, including the $\sim2\%$ difference in median image FWHM, we used identical SExtractors parameters for both fields.  For the detection and deblending parameters, we visually inspected the detected galaxies on the images, and then adjusted the parameters.  Here we aim for a good balance between detecting faint and blended sources and avoiding detecting too many noise spikes, in both deep and shallow regions (Figure~\ref{fig_extraction}). We set the seeing FWHM to be $0\farcs93$, which only affects the star classification output.  Because the image quality is not uniform (Figure~\ref{fig_fwhm_dist}), this value is only an approximate for both fields.  Therefore, star class values in our reference catalogs should have uncertainties somewhat larger than those for narrow-field surveys with uniform image quality. The output catalogs contain both fluxes ($\mu$Jy) and AB magnitudes of galaxies, measured with $1\arcsec$, $1\farcs5$, $2\arcsec$, $3\arcsec$, $4\arcsec$, and $5\arcsec$ diameter apertures, as well as SExtractor's auto apertures, which provide estimates of the total fluxes/magnitudes.  The complete sets of SExtractor input parameters are provided with the data release, so the users can modify them and quickly create their own catalogs that fit their needs.

\subsection{Comparison with Previous Catalogs}

We briefly compare our catalogs with previously published catalogs in COSMOS and SXDS, so the users can be aware of the systematic differences in these datasets.

\begin{figure*}[ht!]
\epsscale{0.7}
\plotone{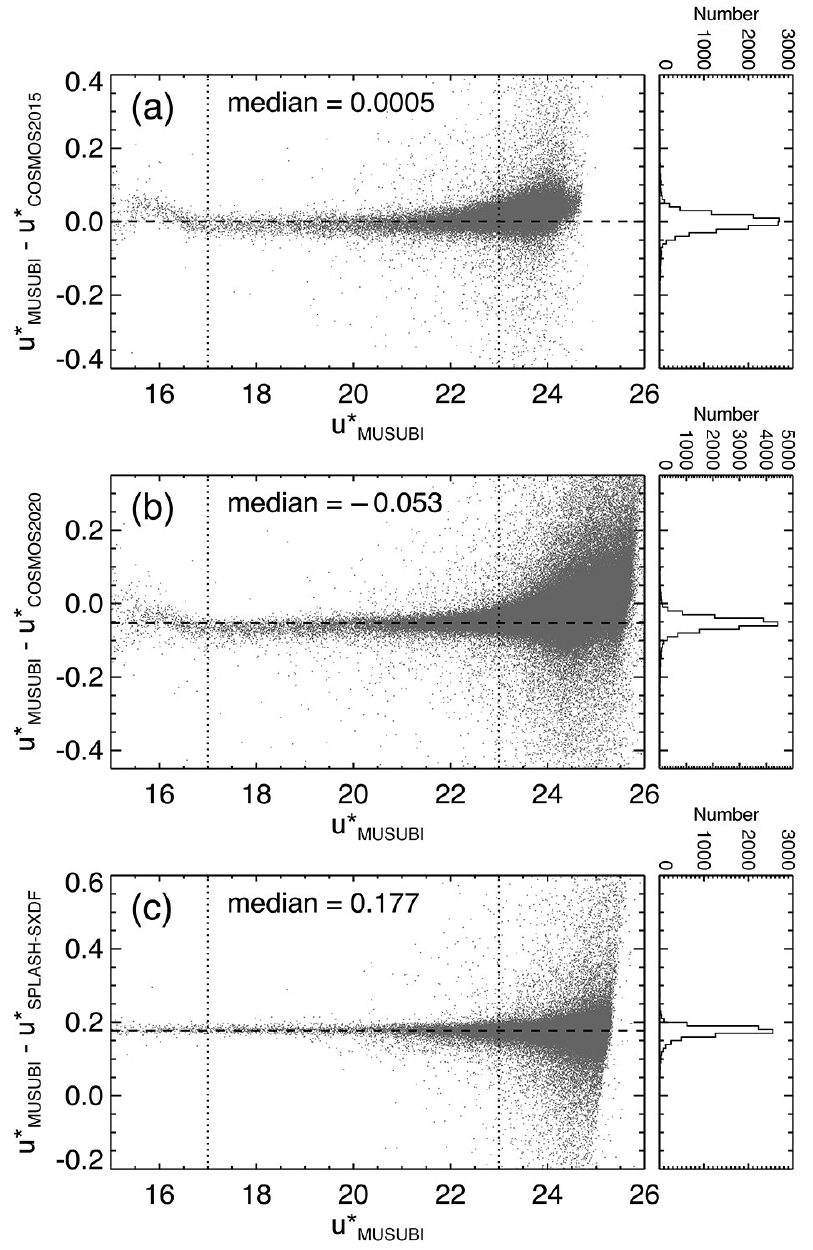}
\caption{Comparisons of photometry between MUSUBI and COSMOS2015 (a), COSMOS2020 (b), and SPLASH-SXDF (c). The comparisons are made on $3\arcsec$ aperture photometry derived from objects with magnitude errors less than 0.05. The median differences for objects with $u^\ast=17$--23 (the two vertical dotted lines) are given in each panel and plotted as the horizontal dashed lines. The histograms show the distributions for $u^\ast=17$--23 objects.
\label{fig_cat_compare}} 
\end{figure*}

\subsubsection{COSMOS2015}
The COSMOS2015 multi-band catalog \citep{laigle16} has been the golden standard for the photometry and photo-$z$ for the COSMOS field.  The CFHT $u^\ast$-band data included in COSMOS2015 were all included in MUSUBI, but MUSUBI also included our new data.  We compared the $3\arcsec$ aperture magnitudes in COSMOS2015 and our reference catalog. This aperture is larger than the optimal aperture for detecting the faintest compact objects.  It is chosen because it is less sensitive to the small PSF size difference between the two datasets.  With this aperture size, the median 5-$\sigma$ limiting magnitudes for COSMOS2015 and MUSUBI are 25.63 and 27.17, respectively.  The numbers of 5-$\sigma$ detected objects are $2.69\times10^5$ and $8.86\times10^5$, respectively.  The area covered by $u^\ast$-band detected objects are 2.62 and 3.25 deg$^2$, respectively.  These differences are mostly caused by the new data obtained after the compilation of COSMOS2015.  In Figure~\ref{fig_cat_compare} (a), we compare the $3\arcsec$ photometry on sources whose magnitude errors are less than 0.05 in both catalogs.  The calibrations of the two catalogs are highly consistent, as the median magnitude difference for objects with $u^\ast = 17$--23 is 0.0005. At the bright end of $u^\ast < 17$, the catalogs start to suffer from saturation effects.

\subsubsection{COSMOS2020}
COSMOS2020 \citep{weaver22} is a new compilation of multi-band data for COSMOS, including Subaru HSC data.
It includes all the available $u^\ast$ data, like MUSUBI, but with its own data reduction.  However, the 5-$\sigma$ limiting magnitude with $3\arcsec$ apertures in COSMOS2020 is 27.11, slightly shallower than ours.  There are $5.90\times10^5$ objects detected at 5 $\sigma$, covering an area of 3.37 deg$^2$.  Their area coverage is comparable to ours, but the detected objects are about 30\% fewer, likely caused by the differences in target selection criteria. In Figure~\ref{fig_cat_compare} (b), we compare the $3\arcsec$ photometry on sources with magnitude errors less than 0.05. There is a $-0.053$ magnitude offset between the two catalogs. This offset should be caused by the different calibration strategies: COSMOS2020 uses SDSS as the calibration reference while we use SNLS.  If we take this offset into account, the difference in limiting magnitudes becomes even larger, i.e., MUSUBI is 0.11 magnitude deeper than COSMOS2020. This should also contribute partially to the larger number of detected objects in the MUSUBI catalog.

\subsection{SPLASH-SXDF}
SPLASH (Spitzer Large Area Survey with Hyper-Suprime-Cam, \citealp{steinhardt14}) is a warm-Spitzer imaging program for both COSMOS and SXDS (aka.\ SXDF).  The multi-band catalog for SPLASH-SXDF was published by \citet{mehta18}, including CFHT $u^\ast$-band photometry from MUSUBI.  Their $u^\ast$-band photometry was derived from an earlier version of our reduced image. The differences between that early version and the present version are the photometric and astrometric references.  So the two catalogs should be highly consistent, in principle. However, the SPLASH-SXDS photometry is derived from images with PSF homogenization across optical, near-IR, and the Spitzer IRAC bands.  As a result, there is not a meaningful direct comparison between the photometry in the catalogs of SPLASH-SXDS and MUSUBI.  In Figure~\ref{fig_cat_compare} (c), we show the comparison between the $3\arcsec$ photometry on sources with magnitude errors less than 0.05 in both catalogs.  There is a 0.177 magnitude offset between the two. If we switch to SExtractor auto-magnitudes, the difference between the two catalogs reduces by $\sim50\%$ but is still quite significant. Such differences are likely caused by the PSF-homogenization process.  On the other hand, it can be seen that the scatter of the differences is much narrower and flatter, comparing to the cases in Figure~\ref{fig_cat_compare} (a) and (b). This reflects the fact that the SPLASH-SXDS and MUSUBI catalogs are derived from images based on identical datasets
and very similar reductions conducted by us.

\section{Application Examples}

\subsection{Photometric Redshifts}\label{photoz}
To demonstrate the value of our MUSUBI $u^\ast$-band data, we compare photo-$z$ derived with and without the $u^\ast$ data. We use the empirical machine learning code, Direct Empirical Photometric Code \citep[DEmP,][]{hsieh14} to derive the photo-$z$. We combined the SExtractor $u^\ast$-band AUTO photometry in the MUSUBI catalog and the HSC second public data release $grizy$ afterburner photometry \citep[HSC PDR2,][]{aihara19} to compile a $u^{\ast}grizy$ multi-band photometric catalog. For the COSMOS field, the training sets are generated by matching the MUSUBI/HSC ${u^\ast}grizy$ multi-band photometric catalog to the redshifts in the COSMOS2020 catalog \citep{weaver22}. The COSMOS2020 catalog has two versions, CLASSIC and FARMER, which are derived using different photometric techniquesc. We used $lp\_ZBEST$ derived using the LePHARE photo-$z$ code \citep{arnouts02,ilb06} in the FARMER catalog as the training references for photo-$z$,. For the SXDS field, we repeated the same procedure to generate the training sets by matching the MUSUBI/HSC multi-band catalog to the redshifts in the SPLASH SXDF catalog \citep{mehta18}. We used $Z\_BEST$ in the SPLASH SXDF catalog as the training references for photo-$z$, which is also derived using LePHARE. We emphasize that the $u^\ast$-band photometry is derived using a different technique from that used to derive the HSC $grizy$ photometry. Therefore any analyses that need accurate $u^\ast - [g,r,i,z,y]$ colors can be seriously affected, such as template SED fitting for photo-$z$. However, because the same photometry/color offsets exist in both the training set and the target set, the conversion between the photometry and the derived quantity (e.g., photo-$z$) should be identical for the training set and the target set. Therefore the effect typically does not impact the results from an empirical machine learning code like DEmP.

Because the training set completely overlaps with the target set, we run the DEmP code in the ``leave-one-out'' mode to prevent overfitting. DEmP always generates a dedicated subset of the training set for each target object. In the leave-one-out mode, DEmP excludes the training object with the identical ID to the target object in the dedicated subset of the training set. With the leave-one-out technique, we are able to compute accurate statistics of the derived quantities (e.g., photo-$z$, or stellar mass, see Section~\ref{sec_green_valley}) for the whole sample. 

\begin{deluxetable*}{lrrrrrrr}[h!]
\tablecolumns{7}
\tablewidth{0pt}
\tablecaption{Photometric Redshift Performance or $u^\ast \leq 27.0$\label{qphotoz}}
\tablehead{
\multicolumn{1}{c}{Samples} &
\multicolumn{3}{c}{ALL} & &
\multicolumn{3}{c}{$z>2.0$} \\
\cline{2-4} \cline{6-8}
& scatter \tablenotemark{a} &
bias \tablenotemark{b} &
f$_{out}$ \tablenotemark{c} & &
scatter \tablenotemark{a} & 
bias \tablenotemark{b} & 
f$_{out}$ \tablenotemark{c}}
\startdata
COSMOS $grizy$ & 0.045 & $-0.0008$ & 19.7\% & & 
0.110 & $-0.059$ & 30.1\% \\
COSMOS${u^\ast}grizy$ & 0.043 & $0.0001$ & 16.8\% & &
0.088 & $-0.050$ & 24.0\% \\
SXDS $grizy$ & 0.067 & $-0.0037$ & 28.2\% & &
0.234 & $-0.132$ & 49.2\% \\
SXDS ${u^\ast}grizy$ & 0.061 & $-0.0016$ & 24.6\% & &
0.153 & $-0.081$ & 40.5\% \\
\enddata
\tablenotetext{a}{$1.48\times$Median Absolute Deviation (MAD) of $\Delta{z}$, 
where $\Delta{z} = \frac{\mathit{photo\mhyphen z} - \mathit{reference\mhyphen z}}{1 + \mathit{reference\mhyphen z}} $}
\tablenotetext{b}{Median of $\Delta{z}$}
\tablenotetext{c}{Outlier fraction: fraction of objects with
$\left | \Delta{z} \right | > 0.15$}
\end{deluxetable*}

\begin{figure*}[ht!]
\begin{center}
  \includegraphics[width=0.9\textwidth,clip]{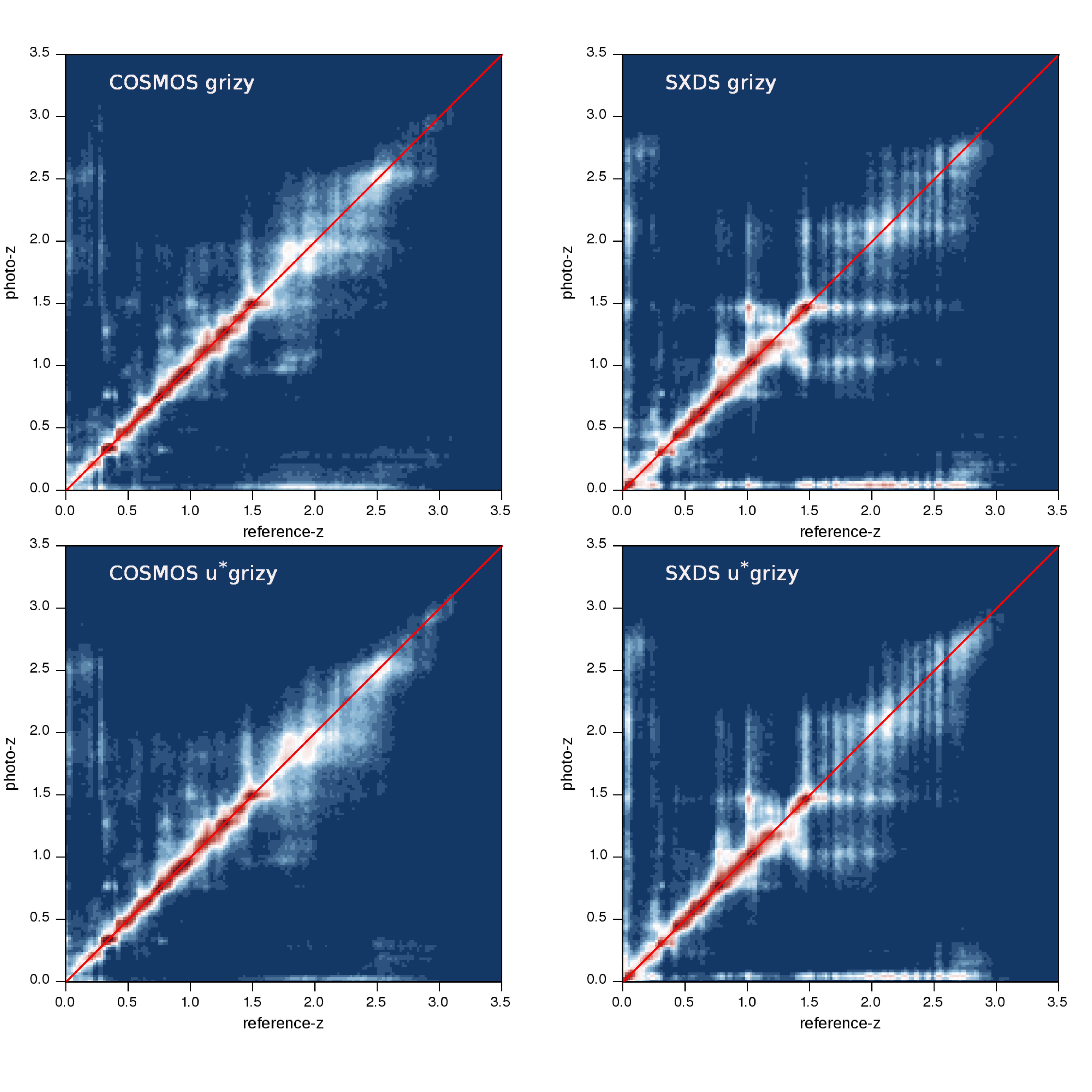}
 \end{center}
\caption{Photometric redshift performance of adding MUSUBI $u^\ast$-band data to HSC $grizy$ data for objects with $u^\ast\leq27.0$. The left panels are for the COSMOS field and the right panels are for the SXDS field. The upper panels are the results derived using the HSC filters alone, while the lower panels are the results derived using the HSC filters and the MUSUBI $u^\ast$-band. Adding the MUSUBI $u^\ast$-band improves the photo-$z$ quality in terms of scatter, bias, as well as outlier fraction.}
\label{fig:photoz}
\end{figure*}

The results are shown in Table~\ref{qphotoz} and Figure~\ref{fig:photoz}. The left panels of Figure~\ref{fig:photoz} are for the COSMOS field, while the right panels are for the SXDS field. All the statistics are calculated for objects with $u^\ast\leq27.0.$ For the COSMOS field, the scatter, bias, and outlier ($\left | \Delta{z} \right | > 0.15$) fraction for the results using the HSC filters alone are 0.045, $-0.0008$, and 19.7\%, respectively. Those with the MUSUBI $u^\ast$-band are 0.043, $0.0001$, and 16.8\%. Adding the $u^\ast$ band improves the scatter, bias, and the outlier fractions. For the SXDS field, the scatter, bias, and outlier fraction for the results using the HSC filters alone are 0.067, $-0.0037$, and 28.2\%, respectively. With the $u^\ast$ band, these values are 0.061, $-0.0016$, and 24.6\%. The scatter, bias, and outlier fraction are all improved by adding the $u^\ast$-band data. These are similar to what we find in the COSMOS field. 

At $z>2$, the photo-$z$ performance improvement extends to the outlier fractions. The photo-$z$ performance mainly relies on strong features of galaxy spectrum such as the Lyman break and the 4000~\AA\ break. The HSC filter with the longest effective wavelength is $y$, which is not able to sample the 4000~\AA\ break for galaxies at $z > 1.44$. Therefore the photo-$z$ scatter increases dramatically at $z > 1.5$. However, adding the MUSUBI $u^\ast$-band data can help sampling the Lyman break ($\lambda_{\rm rest}=1216$~\AA) for galaxies at $z > 2.0$ and the Lyman limit ($\lambda_{\rm rest}=912$~\AA) for galaxies at $z > 3.1$. Therefore, the photo-$z$ performance at $z > 2.0$ can be significantly improved. For galaxies at $z > 2.0$ in the COSMOS field, the photo-$z$ scatter, bias, and outlier fraction are 0.110, $-0.059$, and 30.1\%, respectively when using only the HSC filters. These values are improved to 0.088, $-0.050$, and 24.0\% after adding the $u^\ast$-band data. For the SXDS field, the scatter, bias, and outlier fraction for galaxies at $z > 2.0$ are 0.234, $-0.132$, and 49.2\%, respectively with the HSC filters alone. They are 0.153, $-0.081$, and 40.5\% after adding the $u^\ast$-band data. The improvements in the scatter, bias, and outlier fraction are quite substantial. We conclude that adding $u^\ast$-band data to $grizy$ can significantly improve the photo-$z$ in the HSC UDS fields.

We note that this test is just to demonstrate the improvement of photo-$z$ quality produced by adding the MUSUBI $u^\ast$-band data; the result does not represent the optimal absolute photo-$z$ performance that can be derived using the MUSUBI/HSC catalog.

\subsection{Green-Valley Galaxies at $0.4 < z < 0.6$}\label{sec_green_valley}
We further demonstrate the power of the combination of our $u^\ast$-band data and the HSC UDS $grizy$ data with a mini study of galaxies in the ``green valley'' at $z=0.4$--0.6 down to $10^{9.1}~M_\sun$ for a $u^{\ast} \leq 25$ sample. The green valley is a sparse region between the blue cloud and the red sequence, and is often thought of as the transition zone in which galaxies are in the process of migrating from an active star-forming phase to a quiescence phase \citep{wyd07,mar07,sal14}.  The presence of the low galaxy density in the green valley, if firmly established, can have profound implications in galaxy evolution. For example,  the inferred timescale of galaxy transition can not be very long, otherwise one would expect to see a continuous distribution extended from the green valley to the red color space \citep[e.g.,][]{sal14}. Some other studies, however, have suggested that the quenching time scales of green valley galaxies vary with morphology and environment \citep{sal12,sch14,sme15,jia20}. Therefore, quantifying the fraction of green-valley galaxies and studying their properties provide crucial insights into the quenching mechanisms.

Among various color combinations, the rest-frame $NUV-R$ color has been suggested to be efficient in selecting galaxies with intermediate specific star formation rate between the star-forming and quiescent populations \citep{wyd07,sal14,coe18}. We focus on the redshift of $z\sim0.5$, where the MUSUBI $u^\ast$ band directly samples the rest-frame $NUV$ wavelength, which is sensitive to ongoing star formation. When combined with the HSC UDS $grizy$ photometry, we are able to accurately characterize galaxies in the $NUV-R$ versus stellar mass space and quantify the frequency of galaxies in different populations. \citet{hsieh14} demonstrate that physical quantities besides redshift (e.g., stellar mass) can also be derived from photometry directly. To derive stellar mass and rest-frame $NUV$ and $R$-band luminosity for the green-valley analysis, we repeated the procedure described in Section~\ref{photoz}. For the COSMOS field, the training sets were generated using $lp\_mass\_best$, $lp\_MNUV$, and $lp\_MR$ in the COSMOS2020 FARMER catalog for stellar mass, $NUV$ absolute magnitude, and $R$-band absolute magnitude, respectively. For the SXDS field, $MASS\_BEST$, $LUM\_NUV\_BEST$, and $LUM\_R\_BEST$ in the SPLASH SXDF catalog were used to generate the training sets for stellar mass, $NUV$ luminosity, and $R$-band luminosity, respectively.

We select galaxies with $u^{\ast} \leq 25$. The limiting magnitudes (3\arcsec aperture; 5-$\sigma$) in the HSC 
$grizy$ bands are significantly deeper (27.5, 27.2, 27.0, 26.6, 25.9 in $g, r,i, z,$ and $y$, respectively) than 25.0. Since the observed $u^\ast - [g,r,i,z,y]$ colors are nearly all greater than $-1$ in the redshift range used in this analysis, the $u^{\ast} \leq 25$ selection ensures that the majority of galaxies are also detected in the HSC bands, with non-detection rates in the HSC bands between 0.01\% and 0.05\%.

\begin{figure}[t!]
\epsscale{0.8}
\plotone{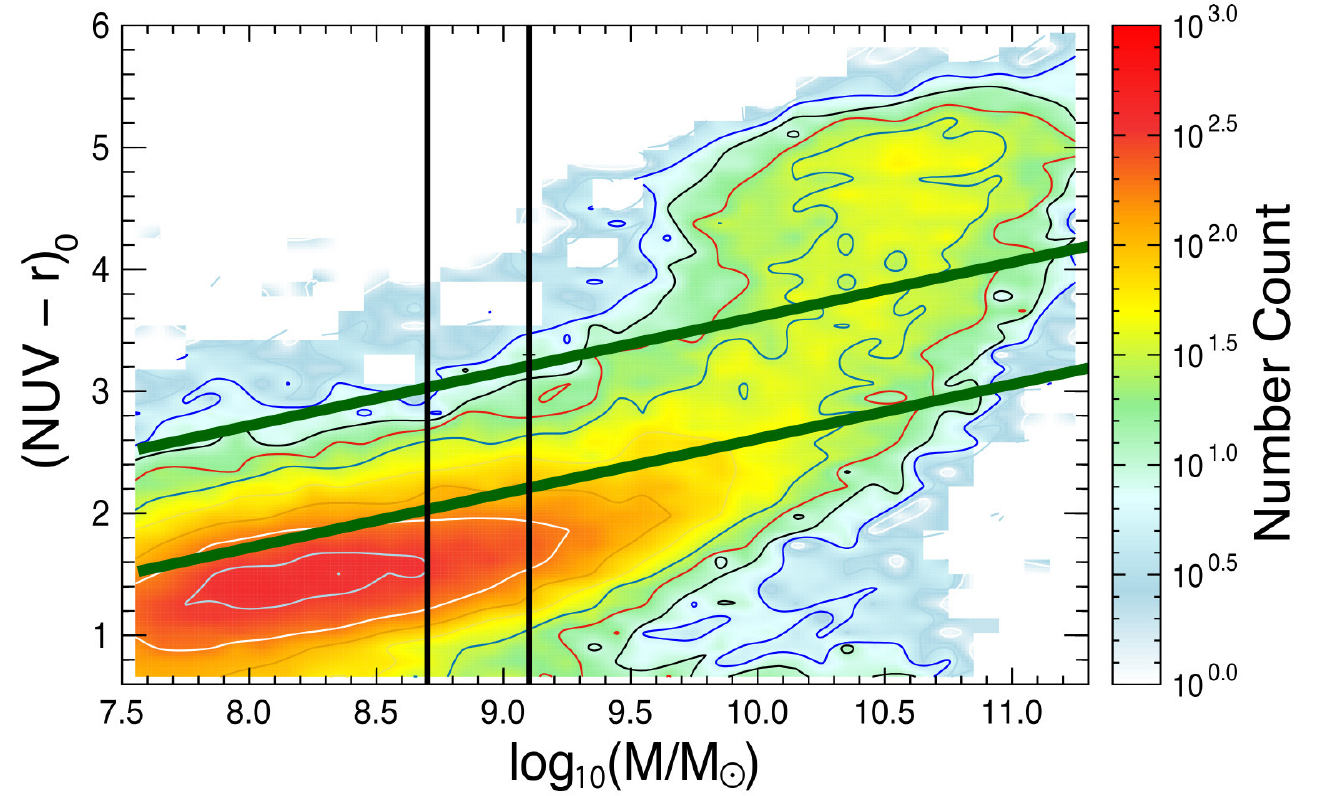}
\caption{Rest-frame $NUV-R$ color vs.\ stellar mass for HSC UDS + MUSUBI galaxies at $0.4 < z < 0.6$. The encoded color scale is in the log scale. The region between two dark green lines is the green valley zone. Solid and dashed black vertical lines denote the mass completeness limits for star-forming and quiescent galaxies, respectively.}
\label{fig:nuvsm}
\end{figure}

\begin{figure}[t!]
\epsscale{0.7}
\plotone{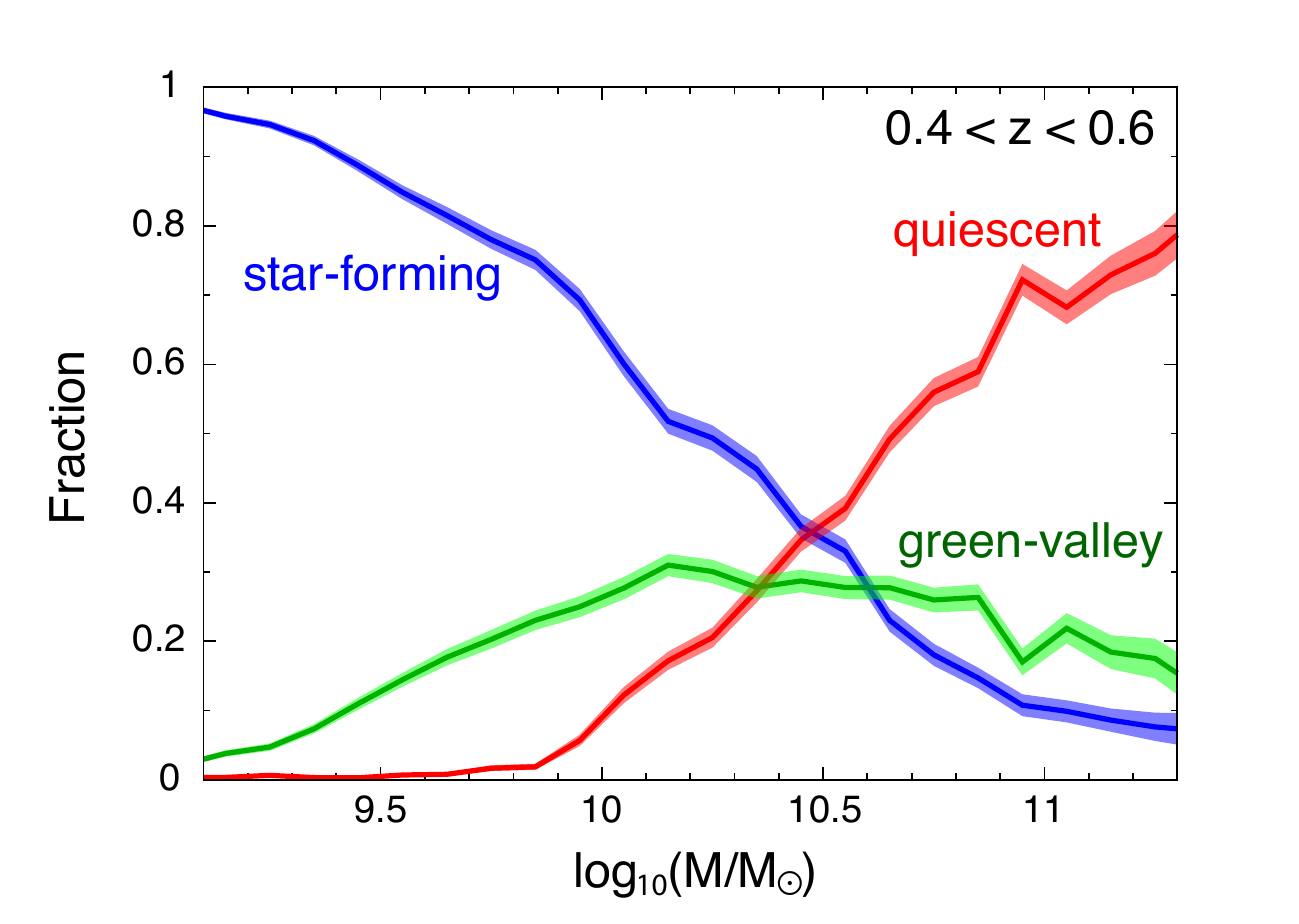}
\caption{Fractions of star-forming (blue), quiescent (red), and green valley galaxies (green) as functions of galaxy stellar mass at $0.4 < z < 0.6$ using the HSC UDS + MUSUBI data. The color bands are the standard deviations estimated using the bootstrap resampling method from 2000 trials. }
\label{fig:gv}
\end{figure}

Figure~\ref{fig:nuvsm} displays the distribution of galaxies of our HSC UDS sample in the redshift range of $0.4 < z < 0.6$. While there are various definitions of the green valley zones in the literature, it may not be straightforward to apply those selections in our dataset directly due to possible systematics in the measurements of color and stellar mass. Therefore, we choose to separate galaxies into three populations, i.e., star-forming, quiescent, and green valley galaxies, by following the iterative procedure similar to the one described in \cite{jia20}. In short, we first adopt a constant $NUV-R = 3.5$ on the $NUV-R$ vs.\ $M_\star$ plane to divide galaxies into two broad groups, blue and red populations, and find the median values of the $NUV-R$ color of the two groups at a given stellar mass bin, separately. We proceed to fit the median $NUV-R$ versus $\log M_\star$ distributions with a linear relation for the two sequences, where the log mass ranges used for the fitting are between 8.7 and 9.5 and between 9.7 and 11.1 for the blue and red populations, respectively. Next, we define the middle points of two sequences as the green valley line and the green valley zone is then defined as the region with $\pm$0.5 color value from the green valley line. The star-forming and quiescent galaxies are referred to as those located above and below the upper and lower boundaries of the green valley, respectively. Once the star-forming and quiescent populations are defined, we fit again the linear relations for the two populations, define the green valley zone, and iterate this process until the green valley converges. The final mass-dependent color criterion for the green valley is as follows:
\begin{equation}\label{eq1}
NUV-R = 0.446 \times \log_{10}(M_\star/M_{\sun}) - 1.348 \pm 0.5.
\end{equation}

In Figure~\ref{fig:gv}, we show the fraction of star-forming (blue), quiescent (red), and green-valley (green) galaxies as functions of $M_\star$, where the sum of the three fractions is unity. The errors are estimated using bootstrap resampling from 2000 runs.  The stellar mass completeness limit is estimated using a method similar to that described in \cite{ilb10}. In the redshift range of $0.4 < z < 0.6$, we compare the stellar mass distributions in $u^{\ast} \leq 25$ and $u^{\ast}\leq 27$ samples, assuming that the $u^{\ast}\leq 27$ sample is complete for the stellar mass range of our interest. We compute the fractions of galaxies with $u^{\ast} \leq 25$ in the complete sample ($u^{\ast}$ $\leq$ 27) as a function of stellar mass. We then define the lower limit of the stellar mass as the mass at which 30$\%$ of the galaxies are fainter than $u^{\ast}$ = 25. Because of the exceptional depth of the HSC and MUSUBI datasets used in this study, the stellar mass completeness limit in our sample reaches down to 10$^{9.1}$ $M_{\sun}$ for quiescent galaxies and 10$^{8.7}$ $M_{\sun}$ for star-forming galaxies, respectively, almost one order of magnitude lower than that in \cite{coe18}. We then choose the mass limit of quiescent galaxies to represent the mass limit for the whole sample to ensure that at this mass limit, star-forming and green-valley galaxies are also complete. It can be seen from Figure \ref{fig:gv} that the fraction of star-forming (quiescent) galaxies is a strong function of stellar mass, decreasing (increasing) rapidly with increasing mass. In contrast, the fraction of green valley galaxies is roughly constant ($\sim$ 25$\%$) in the stellar mass range between 10$^{9.8}$ and 10$^{10.8}$ $M_{\sun}$, but gradually declines to $\sim$0.12 as the stellar mass decreases to 10$^{9.1}$ $M_{\sun}$.  The static green valley fraction at the high mass end is in broad agreement with the results obtained by \citet{coe18}. On the other hand, the small green valley fraction for low-mass galaxies revealed in this study suggests that quenching is inefficient in low mass ($M_\star < 10^{10} M_{\sun}$) galaxies. This result supports the finding by \citet{lin14}, who addressed stellar mass dependent quenching with a different approach and show that the quenching efficiency strongly increases with stellar mass. \citet{lin14} found that the stellar mass quenching becomes dominant over the environmental quenching only for galaxies more massive than $10^{10} M_{\sun}$. We therefore speculate that the low green valley fraction at the low-mass end seen in this work might be due to the lack of stellar mass quenching below $10^{10} M_{\sun}$.

\section{Summary}

We conducted extremely deep $u^\ast$-band imaging with CFHT MegaCam in the COSMOS and SXDS fields, named ``MUSUBI,'' to sample the rest-frame UV of galaxies at $z\lesssim3$ to compliment the Subaru HSC UDS $grizy$ imaging in these two fields. Our deep imaging covers $\gtrsim1$~deg$^2$ in each field. By combining with shallower $u^\ast$ data in the CFHT archive, we reach 5-$\sigma$ limiting magnitudes of $u^\ast=28.1$ and 28.4 on faint galaxies in the deepest areas of our COSMOS and SXDS maps, respectively. In the central 1~deg$^2$ regions, which are more representative for the survey, the limiting magnitudes are 27.7 and 27.8 for COSMOS and SXDS, respectively. The image quality is quite uniform, with FWHM of $0\farcs88$--$0\farcs95$ measured on stars, in the 1~deg$^2$ regions in the two fields. Our photometry is calibrated to the highly accurate CFHT SNLS $u^\ast$ photometry. We estimate that the uncertainties of the calibration are 0.01 magnitude for COSMOS and 0.02 magnitude for SXDS. We tied our astrometry to Gaia DR2. The astrometric uncertainties of our data are 70 mas for COSMOS and 60 mas for SXDS. Using a machine-learning photo-$z$ code, DEmP, we show that adding our $u^\ast$-band data to the HSC $grizy$ data can significantly improve the photo-$z$ in the scatter and bias at $z=0$ to $z\sim3$, and also can mildly improve the photo-$z$ outlier fractions at $z>2$. We also demonstrate that combining the $u^\ast$ and $grizy$ data enables the identification of green-valley galaxies at $z=0.4$--0.6 down to $10^{9.1}~M_\sun$. This allows to study their evolution as a function of stellar mass and their fraction relative to star-forming and quiescent galaxies. We publicly release our reduced and calibrated $u^\ast$ images for COSMOS and SXDS, as well as reference SExtractor catalogs that are science-ready.

\acknowledgments

We thank the referee for comments that greatly improve the manuscript.
We thank the CFHT staff for the observational support, in particular for making the legacy $u^\ast$ filter available to us when MegaCam was migrating to the new filter system. MegaCam is a joint project of CFHT and CEA/DAPNIA, at the Canada-France-Hawaii Telescope (CFHT) which is operated by the National Research Council (NRC) of Canada, the Institut National des Science de l'Univers of the Centre National de la Recherche Scientifique (CNRS) of France, and the University of Hawaii. The observations at the Canada-France-Hawaii Telescope were performed with care and respect from the summit of Maunakea which is a significant cultural and historic site. We are most fortunate to have the opportunity to conduct observations from this mountain. We gratefully acknowledge supports from the Ministry of Science and Technology of Taiwan through grants 110-2112-M-001-006- (W.H.W.), 108-2628-M-001-001-MY3 (L.L.\ and H.Y.J.), and 110-2112-M-001-004 and 109-2112-M-001-005 (Y.T.L.), and from Academia Sinica through the Career Development Awards CDA-107-M03 (L.L.\ and H.Y.J.) and CDA-106-M01 (Y.T.L.). This work was conducted partially when W.H.W. was visiting CFHT as a resident astronomer. W.H.W. am grateful to the hospitality of the CFHT ohana.

\end{document}